\documentclass[fleqn,10pt]{article}
\usepackage{amsthm}
\usepackage{array}
\usepackage{color}
\usepackage{rotating}
\usepackage{eurosym}
\usepackage{cancel}
\usepackage{subfigure}% Support for small, `sub' figures and tables

\date{}

\usepackage{makeidx}
\usepackage{multirow}
\usepackage{multicol}
\usepackage[dvipsnames,svgnames,table]{xcolor}
\usepackage{graphicx}
\usepackage{ulem}
\usepackage{hyperref}
\usepackage{amsmath}
\usepackage{amssymb}
\usepackage{epstopdf}
\usepackage{makecell} 

\begin{document}

\title{ \textbf{Investigating the performance of the Phase II Hotelling $T^2$ chart when monitoring multivariate time series observations} }

\author{Adel Ahmadi Nadi$^{1*}$, Giovanni Celano$^2$, and Stefan H. Steiner$^1$\vspace{0.2cm}\\
{\small {$^1$Department of Statistics and Actuarial Sciences, University of Waterloo, Waterloo, Canada}}\\
\small {$^2$Department of Civil Engineering and Architecture, University of Catania, Catania, Italy}\\
\small {$^*$Corresponding author email: adel.nadi67@gmail.com}}

\maketitle 

\begin{abstract}
Thanks to high-tech measurement systems like sensors, data are often collected with high frequency in modern industrial processes. This phenomenon could potentially produce autocorrelated and cross-correlated measurements. It has been shown that if this issue is not properly accounted for while designing the control charts, many false alarms may be observed resulting in disruption of the monitoring process efficiency. There are generally two recommended ways to monitor autocorrelated data: fitting a time series model and then monitoring the residuals or directly monitoring the original observations. Although residual charts are popular in the literature because they offer advantages such as ease of implementation, questions have been raised about their efficiency due to the loss of information. This paper develops the required methodology for applying Hotelling's $T^2$ chart directly to monitoring multivariate autocorrelated/cross-correlated observations. To model such data, we use the multivariate vector autoregressive time series model of order $p \geq 1$, say VAR($p$). We compare the performance of the $T^2$ chart based on the original observations and the residual-based $T^2$ chart through the well-known metric average run length and a newly introduced criterion called first-to-signal. The study's results indicate that the proposed method consistently outperforms the alternative chart. We also illustrate the proposed method using two examples, one example from a steel sheet rolling process (VAR(1) observations) and one from a chemical process (VAR(3) observations).
 \end{abstract}

%\noindent AMS subject classifications: 62N05, 90B25, 60K10.

\section{Introduction}

Statistical process monitoring (SPM) involves the systematic application of statistical methods to track and analyze the performance of a process over time. In particular, the main focus of SPM is to develop efficient control charts as graphical tools to monitor the quality of a process and to improve its performance by providing valuable insights into process stability and variability. Though SPM was first introduced to monitor industrial processes (e.g., a manufacturing process), it is now also applied in many nonindustrial sectors including medicine and healthcare surveillance (Steiner et al. (2000)), financial markets (Bisiotis et al. (2022)), environment (Xi and Qiu, (2023)), network monitoring (Wilson et al. (2019)), and change-point problems (Wang and Zwetsloot (2024)). Traditionally, the well-known $\overline{X}$-chart and $S^2$-chart are the most popular choices for assessing process stability. These control charts are advantageous in practice due to their ease of implementation and interpretation. However, their performance heavily depends on certain assumptions: (i) the process output follows a normal or near-normal distribution, (ii) the process quality being monitored is described by a single characteristic, and (iii) the observations are independent. However, these assumptions might be violated in many real applications, especially due to recent advances in data acquisition technologies. As Perry and Wang (2022) noted, the typical assumptions required by traditional control charting strategies are not likely to be met by today’s more modern processes. Accordingly, advanced monitoring procedures are required for modern quality engineering practices to avoid the negative effects of these assumption violations.

Nowadays, thanks to the evolution of measurement equipment, several quality characteristics are often measured simultaneously. On the other hand, high-tech sensor data collectors have permitted the rapid collection of data over time at a relatively low cost (Grimshaw (2023)). This typically results in highly autocorrelated and potentially cross-correlated consecutive observations. Prajapati and Singh (2012) stated that autocorrelation may exist among observations in different processes due to several reasons:
\begin{itemize}
\item[•] Manufacturing processes: Due to computer control of production equipment and downstream pooling of multiple stream processes.
\item[•] Chemical process: Due to inertia of large batches, continuous flow of material and feedback and feed-forward process control systems. 
\item[•] Service processes: As described in queuing theory, waiting times for customers are often influenced by the waiting time of the previous customers.
\end{itemize}

In addition to the above-discussed processes, there are many other types of industrial and non-industrial processes like the financial processes (Sadeghi et al. (2017)), network monitoring (Taheri et al. (2020)), agricultural crop production process  (Mohammed, Z. M. (2022)), pharmaceutical industry  (Prajapati and Singh (2016)),  construction (Barbeito et al. (2017)), textile processes (Keshavarz et al. (2019)), air pollution surveillance (Xie and Qiu (2022, 2023)), where specific correlation patterns can be observed among observations.  

    Hotelling's (1947) $T^2$ control chart is perhaps the most well-known method for monitoring the mean vector of a multivariate normal process. Numerous studies in the SPM literature have explored the performance of various multivariate control charts, including the $T^2$ chart, in the presence of autocorrelation and cross-correlation. There exist two main approaches for the statistical monitoring of autocorrelated data. The first approach is to fit a class of multivariate time series models to the data and then monitor the resulting residuals. These charts are also known as residual charts. Multivariate time series models like the vector autoregressive (VAR) and the vector autoregressive-moving-average (VARMA) are efficient statistical techniques to account for autocorrelation and cross-correlation among observations. For example, Jarrett and Pan (2007) proposed to use Hotelling's $T^2$ for monitoring the residuals generated by an autocorrelated VAR model. Pan and Jarrett (2012) also comprehensively investigated the performance of a residual-based Hotelling's $T^2$ control chart under a VAR model of order $p \geq 1$, say VAR$(p)$. Huang et al. (2014) and Xu and Li (2021) also investigated a residual-based $T^2$ control chart for monitoring a VARMA($p,q$) process where $p \geq 1$ is the autoregressive order and $q \geq 1$ is the moving-average order. Recently,  Lee and Lee (2023) developed control charts to monitor VAR($p$) and structural VAR($p$) time series data using the residual-based cumulative sum technique.

 Although popular in the literature, many questions have been raised about the residual charts, for example, about their interpretability, insufficiency in detecting some assignable causes, efficiency, the effect of model misspecification, dependence among the residuals and loss of information (Yao et al. 2023). The second approach for monitoring multivariate time series data is to monitor the original observations directly by assuming a specific time series model (e.g., the VARMA model) for the data and paying particular attention to the variance-covariance structure while constructing the control limits. The time series models considered in this approach are simpler than those considered in the residual-based control chart.  One of the most important studies in this area was conducted by Kalgonda and Kulkarni (2004), who developed a $Z$ chart for monitoring the mean vector of multivariate autocorrelated processes under a VAR(1) model. Through some numerical studies based on a reduced VAR(1) model, i.e., a model with no cross-correlation, Kalgonda and Kulkarni (2004) showed that autocorrelation has a serious impact on the performance of conventional control charts for multivariate data. More recently, Leoni et al. (2015c)  proposed two monitoring schemes, simultaneous univariate (SU) $\overline{X}$ charts and Hotelling's $T^2$ control chart, when the within-sample observations in each inspection follow a reduced bivariate VAR(1) model, but the chart's statistics of consecutive inspections are independent. Under the same assumptions, Leoni et al. (2015b)  studied the effect of autocorrelation on the performance of the $T^2$ chart when the observations are described by a reduced VAR (1) time series model. A synthetic chart to monitor bivariate processes with autocorrelated data was designed by Leoni et al. (2015a)  based on the reduced VAR(1) model. Leoni et al. (2016) suggested a new sampling strategy to reduce the negative effect of the autocorrelation on the performance of the $T^2$ chart when monitoring the mean vector of a bivariate normal process with VAR(1) observations.  For the same process as described in Leoni et al. (2015c), Simões et al. (2016) introduced Synthetic SU $\overline{X}$ and Hotelling's $T^2$ control charts. Recently, Sabahno et al. (2021) extended the $T^2$ control chart applications to a VARMA(1,1) process. 

Although some researchers have developed process monitoring approaches based on VARMA models, authors like Pan and Jarrett (2012) and Snoussi (2011) listed several advantages of the VAR models over the VARMA models in SPM practice. For example, Pan and Jarrett (2012) demonstrated that the significance of moving average terms in multivariate time series models is reduced as disturbance effects are absorbed by autoregressive components, making MA terms less critical compared to univariate models. The other important issue is that the number of parameters needed to be estimated in the VARMA($p,q$) model quickly becomes overwhelmingly large with increasing orders of $p$ and $q$ and can cause estimation issues during the model fitting stage.  Furthermore, estimating a VAR model is much easier than estimating a VARMA model. Moreover, some other researchers demonstrated that autoregressive models are most frequently encountered in practice (Montgomery and Mastrangelo (1991) and Alwan and Roberts (1988)). Accordingly, this study focuses on VAR models.

Having reviewed the relevant literature, we found that, to the best of our knowledge, no studies in the SPM field utilize the original observations from a VAR($p$) model, for any order $p \geq 1$, to monitor the process mean vector. By calculating the closed-form variance-covariance matrix of the sample average vector, our method leverages the full association-related information (correlation, autocorrelation, and cross-correlation) contained within the time series data. This approach enhances the detection power of the $T^2$ chart and addresses the challenges associated with residual-based methods. Motivated by the existing gap in the literature, the main aim of this study is to extend the results of Leoni et al. (2015a), Leoni et al. (2015b), and Leoni et al. (2015c) in the following directions:
\begin{itemize}
\item[•] \textit{Number of quality characteristics:} from the bivariate to the multivariate scenario.
\item[•] \textit{The VAR(1) model:} from the reduced VAR(1) model to a full VAR(1) model by considering both autocorrelation and cross-correlation.
\item[•] \textit{The order of VAR model:} from the VAR(1) model to the general class of VAR($p$) models.
\end{itemize}

By doing this, we only consider the Phase II implementation of the Hotelling's $T^2$ control chart. This paper is organized as follows.  Section 2 introduces the VAR($p$) model and its subclass VAR($1$) as well as the way that a VAR($p$) model can be represented as a VAR($1$) model. The required methodology to calculate Hotelling's $T^2$ chart's monitor statistic and its associated control limit is discussed in Section 3. Section 4 presents the results of an extensive numerical analysis conducted to evaluate the performance of the chart and compare it with existing methods. The proposed approach is illustrated using two examples in Section 5, one example from a steel sheet rolling process (VAR(1) observations) and one from a chemical process (VAR(3) observations).  Finally, the paper concludes in Section 6 with a concise summary of the findings and conclusions as well as some directions for further research.

\section{The Vector Autoregressive VAR($p$) model}
\label{sec:Var(1)} 

Let us consider a process producing items where the quality of each item is measured by $v$ correlated continuous quality characteristics. We denote $\textbf{X}_t=(X_{t,1},X_{t,2},\cdots,X_{t,v})^{\intercal}$ the vector of random characteristics at time $t$, where the superscript $\intercal$ stands for transpose. The vector $\textbf{X}_t$ is assumed to follow a $v$-variate normal distribution with mean vector $\pmb{\mu}_{\textbf{X}}$ and variance-covariance matrix $\pmb{\Sigma}_{\textbf{X}}$; i.e.,  $\textbf{X}_t \sim \textbf{MN}_{v}(\pmb{\mu}_{\textbf{X}},\pmb{\Sigma}_{\textbf{X}})$. Due to potentially high-frequency data collection, observations collected at consecutive inspections are assumed to be correlated (and also cross-correlated). This means that the  vector $\mathbf{X}_{t}$ could depend on the observations collected during the last $p \geq 1$ inspections $\mathbf{X}_{t-1}, \mathbf{X}_{t-2},\cdots,\mathbf{X}_{t-p-1}$ and $\mathbf{X}_{t-p}$. To describe this relationship between $\mathbf{X}_{t}$ and the data obtained from its preceding inspections, we applied the following multivariate vector autoregressive time series model of order $p$, denoted by VAR($p$), as
\begin{align}\label{equ:varp}
\textbf{X}_{t}-\pmb{\mu}_{\textbf{X}}=\pmb{\Phi}_{1}(\textbf{X}_{t-1}-\pmb{\mu}_{\textbf{X}})+\cdots+\pmb{\Phi}_{p}(\textbf{X}_{t-p}-\pmb{\mu}_{\textbf{X}})+\pmb{\epsilon}_{t}, \, t=0,\pm 1, \pm 2, \cdots,
\end{align}
where $\pmb{\Phi}_{i}$ for $i=1,\cdots,p$ is the $v \times v$ coefficients matrix such that $\pmb{\Phi}_{p} \neq 0$ and $\pmb{\epsilon}_t$ are independent $v$-variate normal vectors with mean zero and covariance matrix $\pmb{\Sigma}_{\pmb{\epsilon}}$, i.e., $\pmb{\epsilon}_t \sim \textbf{MN}_v(\textbf{0},\pmb{\Sigma}_{\pmb{\epsilon}})$. We consider that the mean vector $\pmb{\mu}_{\textbf{X}}$, the coefficients matrices $\pmb{\Phi}_{1},\ldots,\pmb{\Phi}_{p}$, and the error vector variance-covariance matrix $\pmb{\Sigma}_{\epsilon}$ are estimated from a sufficiently large data set taken from a stable process during the Phase I implementation of the control chart. Thus, hereafter, we assume that all these parameters are \emph{known}. With a VAR($p$) model, each individual quality characteristic measured at time $t$ (each element of the vector $\mathbf{X}_{t}$) not only depends on its observations collected at times $t-1,t-2,\cdots,t-p$ (autocorrelation) but can also depend on the observations collected for the other $v-1$ quality characteristic at times $t-1,t-2,\cdots,t-p$, (cross-correlation). The VAR($p$) model can also be represented in the following form 
\begin{align}\label{equ:varpC}
\textbf{X}_{t}=\textbf{C}+\pmb{\Phi}_{1}\textbf{X}_{t-1}+\cdots+\pmb{\Phi}_{p}\textbf{X}_{t-p}+\pmb{\epsilon}_{t}, \, t=0,\pm 1, \pm 2, \cdots,
\end{align}
where $\textbf{C}$ is the vector of constants. If we set $\textbf{C}=(\mathbf{I}_{v}-\pmb{\Phi}_{1}-\ldots-\pmb{\Phi}_{p})\pmb{\mu}_{\textbf{X}}$, where $\mathbf{I}_{v}$ is the $v\times v $ identity matrix, then both models in \eqref{equ:varp} and \eqref{equ:varpC} are identical.  For the multivariate time series $\textbf{X}_t$, the lag $k$ cross-covariance matrix can be obtained as
\begin{align}\label{equ:attcovmat}
\pmb{\Gamma}(k)&=\mathbb{E}\big[(\textbf{X}_{t}-\pmb{\mu}_{\textbf{X}})(\textbf{X}_{t-k}-\pmb{\mu}_{\textbf{X}})^{\intercal}\big]=\begin{pmatrix}\gamma_{j,h}(k)\end{pmatrix}_{v\times v},
\end{align}
for $k=0,\pm1,\pm2,\pm3,...,$ and $j,h=1,\ldots,v$, where $\gamma_{j,h}(k)=\mathbb{E}\left[(X_{t,j}-\mu_{j})(X_{t-k,h}-\mu_{h})\right]$ is the covariance between $X_{t,j}$ (the quality characteristic $j$ measured at time $t$) and $X_{t-k,h}$ (the quality characteristic $h$ measured at time $t-k$) for $j,h=1,2,\cdots,v$. Thus, $\pmb{\Gamma}(k)$ measures the linear dynamic dependence between the quality characteristics of consecutive inspections. By definition, the variance-covariance matrix of $\textbf{X}_t$ is $\pmb{\Sigma}_{\textbf{X}}=\pmb{\Gamma}(0)$. We assume that the VAR($p$) model in \eqref{equ:varp} is (weakly) stationary, which means that its mean vector $\pmb{\mu}_{\textbf{X}}$ and cross-covariance matrix $\pmb{\Gamma}(k)$ are time-invariant; i.e. they are not functions of time $t$. It has been shown that the necessary and sufficient conditions for a VAR(p) model to be stationary is that the solutions of the equation $|\mathbf{I}_{v}-\pmb{\Phi}_{1} B -  \ldots-\pmb{\Phi}_{p} B^p|=0$ are greater than 1 in absolute value where $|\textbf{A}|$ stands for the determinant of matrix $\textbf{A}$ and $B$ is the backshift operator (see Tsay 2013).

 To derive the cross-covariance matrix $\pmb{\Gamma}(k)$ of a Var($p$) model, one can use the following recursive relation in addition to the general relation \eqref{equ:attcovmat} (see Tsay 2013)
\begin{align}\label{equ:attcovmatvarp}
\pmb{\Gamma}(k)=\pmb{\Phi}_{1}\pmb{\Gamma}(k-1)+\pmb{\Phi}_{2}\pmb{\Gamma}(k-2)+\cdots+\pmb{\Phi}_{p}\pmb{\Gamma}(k-p), \, \text{for}\, k>0.
\end{align} 

Relation \eqref{equ:attcovmatvarp} and the fact that $\pmb{\Gamma}(-k)=\pmb{\Gamma}(k)^{\intercal}$ can be used to derive $\pmb{\Gamma}(k)$ when $k<0$. Moreover, to calculate the variance-covariance matrix of time series $\textbf{X}_{t}$   in \eqref{equ:varp}, we have 
\begin{align}\label{equ:sigmavarp}
\pmb{\Sigma}_{\textbf{X}}=\pmb{\Gamma}(0)=\pmb{\Phi}_{1}\pmb{\Gamma}(-1)+\pmb{\Phi}_{2}\pmb{\Gamma}(-2)+\cdots+\pmb{\Phi}_{p}\pmb{\Gamma}(-p)+\boldsymbol{\Sigma}_{\pmb{\epsilon}}.
\end{align} 

Relation \eqref{equ:sigmavarp} states that we need to determine the matrices $\pmb{\Gamma}(-1),\ldots,\pmb{\Gamma}(-p)$ to be able to calculate $\pmb{\Sigma}_{\textbf{X}}$. Throughout the next two subsections, we provide the required methodology not only to calculate $\pmb{\Sigma}_{\textbf{X}}$ in a much simpler way only based on the given matrices $\pmb{\Phi}_{1},\ldots,\pmb{\Phi}_{p}$ and $\pmb{\Sigma}_{\pmb{\epsilon}}$, but also to derive the variance-covariance matrix $\pmb{\Sigma}_{\overline{\textbf{X}}}$ of the vector of sample means $\overline{\textbf{X}}_t$ at time $t$.

\subsection{The VAR($1$) model}
\label{sec:var1}
The VAR(1) model, i.e., the VAR$(p)$ model with $p=1$, can be used to capture temporal short-ranged interdependencies among multiple quality variables at each time as well as the dependencies between  $\mathbf{X}_{t}$ and $\mathbf{X}_{t-1}$.  According to this model, $\mathbf{X}_{t}$ and $\mathbf{X}_{t-1}$ could be autocorrelated and cross-correlation according to the following matrix notation
\begin{equation}
 \label{equ:VAR1}
 \mathbf{X}_{t}-\pmb{\mu}_{\mathbf X}=\pmb{\Phi}(\mathbf{X}_{t-1}
-\pmb{\mu}_{\mathbf X})+\pmb{\epsilon}_{t}, \quad t=0,\pm 1, \pm 2,...,
\end{equation}
where
\begin{align}\label{equ:meanCov}
\pmb{\mu}_{\mathbf X}=(\mu_{X_1},\cdots,\mu_{X_v})^{\intercal},\qquad \text{and} \qquad \boldsymbol{\Phi}=\begin{pmatrix}
 \Phi_{X_{1}X_{1}} & \Phi_{X_{1}X_{2}}& \cdots & \Phi_{X_{1}X_{v}} \\
\Phi_{X_{2}X_{1}} & \Phi_{X_{2}X_{2}}& \cdots & \Phi_{X_{2}X_{v}} \\ 
\vdots &\vdots &\ddots &\vdots\\
 \Phi_{X_{v}X_{1}} & \Phi_{X_{v}X_{2}}& \cdots & \Phi_{X_{v}X_{v}} \\ 
\end{pmatrix}_{v \times v},
\end{align}
are respectively the mean vector of $\mathbf{X}_{t}$ and $v\times v$ coefficients matrix and $\pmb{\epsilon}_{t}=(\epsilon_{X_{t,1}},\ldots,\epsilon_{X_{t,v}})^{\intercal} \sim \textbf{MN}_{v} \left(\textbf{0},\pmb{\Sigma}_{\pmb{\epsilon}}\right)$ with
\begin{align}\label{equ:SigmaE}
\pmb{\Sigma}_{\pmb{\epsilon}}=\begin{pmatrix}
 \sigma_{e_{X_1}}^2 & \sigma_{e_{X_{1}X_{2}}}&\cdots & \sigma_{e_{X_{1}X_{v}}}\\
 \sigma_{e_{X_{2}X_{1}}} & \sigma_{e_{X_2}}^2&\cdots & \sigma_{e_{X_{2}X_{v}}}\\
\vdots &\vdots &\ddots &\vdots\\
\sigma_{e_{X_{v}X_{1}}} & \sigma_{e_{X_{v}X_{2}}} &\cdots & \sigma_{e_{X_v}}^2\\
\end{pmatrix}_{v \times v}.
\end{align}  

After some mathematical manipulations, relation \eqref{equ:VAR1} can be represented by the following system of $v$ equations
\begin{align}\label{equ:var(1)}
 X_{t,1}&=C_1+\Phi_{X_{1}X_{1}}X_{t-1,1}+\cdots + \Phi_{X_{1}X_{v}} X_{t-1,v}+\epsilon_{X_{t,1}}\nonumber\\
\vdots \nonumber\\
 X_{t,v}&=C_v+\Phi_{X_{v}X_{1}}X_{t-1,1}+\cdots + \Phi_{X_{v}X_{v}} X_{t-1,v}+\epsilon_{X_{t,v}},
\end{align}
where $C_j=\mu_{X_j}-\sum_{h=1}^v \Phi_{X_{i}X_{h}}  \mu_{X_j}$ for $j=1,2,\cdots,v$. The relation \eqref{equ:var(1)} helps us to better understand the relationship between different quality variables coming from a VAR model.  In particular, relation \eqref{equ:var(1)} demonstrates that the VAR($1$) model not only accounts for the autocorrelation between $X_{t,j}$ and $X_{t-1,j}$ ($j=1,2,\cdots,v$)  through the coefficient $\Phi_{X_{j}X_{j}}$, but it also takes into account the cross-correlation between $X_{t,j}$ and $X_{t-1,h}$ via the coefficient $\Phi_{X_{j}X_{h}}$ ($j\neq h,\, j,h=1,2,\cdots,v$). For instance, $\Phi_{X_{1}X_{2}}$ expresses the cross-correlation between $X_{t,1}$ and $X_{t-1,2}$ in the presence of $X_{t-1,1},X_{t-1,3},\cdots,X_{t-1,v}$. On the other hand, if the off-diagonal elements of $\boldsymbol{\Phi}$ in \eqref{equ:meanCov} are zero, then the $X_{t,j}$'s for $j=1,2,\cdots,v$ are not cross-correlated and can be modelled using univariate autoregressive models AR(1). But note that these AR(1) models are still instantaneously (cross-sectionally) correlated. If $\boldsymbol{\Phi}$ in \eqref{equ:meanCov} and $\pmb{\Sigma}_{\pmb{\epsilon}}$ in \eqref{equ:SigmaE} are both diagonal matrices, then the $X_{t,j}$'s for $j=1,2,\cdots,v$ could be modeled using independent AR(1) models. 

It is worth mentioning that the stationarity of a VAR(1) model can also be assessed based on the features of matrix $\pmb{\Phi}$. The necessary and sufficient condition for a VAR(1) to be stationary is that all eigenvalues of $\pmb{\Phi}$ must be less than 1 in absolute value.  The variance-covariance matrix $\pmb{\Sigma}_{\textbf{X}}$ of a VAR(1) model can be derived as a function of matrices $\pmb{\Phi}$ and $\pmb{\Sigma}_{\pmb{\epsilon}}$ using the following relation (see Tsay 2013)
\begin{align} \label{equ:VecSigmaW}
 \mathrm{Vec}(\pmb{\Sigma}_{\textbf{X}})&=(\sigma^{2}_{X_1},\cdots, \gamma_{1,v}(0),\gamma_{2,1}(0),\cdots,\gamma_{2,v}(0),\cdots,\gamma_{v,1}(0),\cdots,\sigma^2_{X_v})_{v^2,1}^{\intercal}\nonumber\\
&=(\mathbf{I}_{v^2}-\pmb{\Phi}\otimes\pmb{\Phi})^{-1}
 \mathrm{Vec}(\pmb{\Sigma}_{\pmb{\epsilon}}),
\end{align}
where  $\otimes$ stands for the Kronecker product and $\mathrm{Vec}$ is the operator that transforms a matrix into a one-column vector by stacking its columns. In addition, post multiplying Equation \eqref{equ:VAR1} by $(\textbf{X}_{t-k}-\pmb{\mu}_{\textbf{X}})$ and taking expectation, we have 
\begin{align}\label{equ:gammak}
\pmb{\Gamma}(k)=\pmb{\Phi}\pmb{\Gamma}(k-1)=\pmb{\Phi}^2\pmb{\Gamma}(k-2)=\cdots=\pmb{\Phi}^k\pmb{\Sigma}_{\textbf{X}}.
\end{align}

Now, we are in a position to derive the variance-covariance matrix of the random vector of sample means which will be used in the next section to define the Hotelling's $T^2$ chart's monitor statistic. Let us denote by $\textbf{X}_{t,1},....,\textbf{X}_{t,n}$ the random sample of size $n$  taken at time $t$ from the process. Moreover, let $\overline{\textbf{X}}_t=(\overline{X}_{t,1},\cdots,\overline{X}_{t,v})^{\intercal}$ be the vector of sample means at time $t$ so that $\overline{X}_{t,j}=\frac{1}{n}\sum_{i=1}^n X_{t,j,i}$ where $X_{t,j,i}$ is the observation $i=1,2,...,n$ of the quality characteristic $j=1,\ldots,v$ in the sample collected at time $t$.  Given stationary multivariate time series $\textbf{X}_t$ with the cross-covariance matrix  $\pmb{\Gamma}(k)$ at lag $k$, Reinsel (2003) showed that $\overline{\textbf{X}}_t \sim \textbf{MN}_v(\pmb{\mu}_{\overline{\textbf{X}}},\pmb{\Sigma}_{\overline{\textbf{X}}})$ where
\begin{align}\label{equ:SigmaXbar}
 \pmb{\mu}_{\overline{\textbf{X}}}=\pmb{\mu}_{\textbf{X}}, \quad \pmb{\Sigma}_{\overline{\textbf{X}}}=\frac{1}{n^2}\sum_{l=1}^{n}\sum_{s=1}^{n}\pmb{\Gamma}(l-s)=\frac{1}{n^2}\sum_{k=-(n-1)}^{n-1}\left(n-|k|\right)\pmb{\Gamma}(k).
\end{align}

To derive the variance-covariance matrix $\pmb{\Sigma}_{\overline{\textbf{X}}}$ of a VAR(1) model, we applied equation \eqref{equ:SigmaXbar} and obtained
 \begin{align}\label{equ:A3VCwbar}
 \pmb{\Sigma}_{\overline{\textbf{X}}}&=\frac{1}{n}\pmb{\Sigma}_{\textbf{X}}+\frac{1}{n^2}\sum_{k=1}^{n-1}(n-k)\pmb{\Gamma}(k)+\frac{1}{n^2}\sum_{k=1}^{n-1}(n-k)\pmb{\Gamma}(-k) \nonumber\\
 &=\frac{1}{n}\pmb{\Sigma}_{\textbf{X}}+\frac{1}{n^2}\sum_{k=1}^{n-1}(n-k)[\pmb{\Gamma}(k)+\pmb{\Gamma}(k)^{\intercal}] \nonumber\\
 &=\frac{1}{n}\pmb{\Sigma}_{\textbf{X}}+\frac{1}{n^2} \sum_{k=1}^{n-1}(n-k) \Big[\pmb{\Phi}^{k}\pmb{\Sigma}_{\textbf{X}}+\pmb{\Sigma}_{\textbf{X}} \pmb{\Phi}^{\intercal^{k}} \Big] 
 \nonumber\\
 &=\frac{1}{n}\pmb{\Sigma}_{\textbf{X}}+\frac{1}{n}\sum_{k=1}^{n-1}\Big[\pmb{\Phi}^{k}\pmb{\Sigma}_{\textbf{X}}+\pmb{\Sigma}_{\textbf{X}} \pmb{\Phi}^{\intercal^{k}} \Big] -\frac{1}{n^2} \sum_{k=1}^{n-1} k\Big[\pmb{\Phi}^{k}\pmb{\Sigma}_{\textbf{X}}+\pmb{\Sigma}_{\textbf{X}} \pmb{\Phi}^{\intercal^{k}} \Big] ,
 \end{align} 
where $\pmb{\Sigma}_{\textbf{X}}$ is given in \eqref{equ:VecSigmaW}, the second equality is derived using the fact that  $\pmb{\Gamma}(-k)=\pmb{\Gamma}(k)^{\intercal}$, and the third equality is obtained based on \eqref{equ:gammak}. We further simplify $ \pmb{\Sigma}_{\overline{\textbf{X}}}$ in \eqref{equ:A3VCwbar} for calculation convenience.  Using Proposition 1.5.38. of Hubbard and Hubbard (2015), we can show that $\sum_{k=1}^{n-1} \boldsymbol{\Phi}^{k}=\Lambda(\boldsymbol{\Phi})$ and $\sum_{k=1}^{n-1} k\boldsymbol{\Phi}^{k}=\Pi(\boldsymbol{\Phi})$ where 
\begin{align}\label{equ:HH}
 \Lambda(\boldsymbol{\Phi})&=(\boldsymbol{\Phi}-\boldsymbol{\Phi}^{n})(\mathbf{I}_{v}-\boldsymbol{\Phi})^{-1}\nonumber\\
 \Pi(\boldsymbol{\Phi})&=(\boldsymbol{\Phi}^{-1}-\mathbf{I}_{v})^{-1}\Big((\mathbf{I}_{v}-\boldsymbol{\Phi}^{n-1})(\mathbf{I}_{v}-\boldsymbol{\Phi})^{-1}-(n-1)\boldsymbol{\Phi}^{n-1}\Big).
\end{align}

To derive $\Lambda(\boldsymbol{\Phi})$ and $ \Pi(\boldsymbol{\Phi})$ in \eqref{equ:HH}, we need the absolute values of all eigenvalues of $\pmb{\Phi}$ to be less than 1, which we have already as the stationarity condition of the VAR(1) model. Rewriting relation \eqref{equ:A3VCwbar} based on $\Lambda(\boldsymbol{\Phi})$ and $\Pi(\boldsymbol{\Phi})$, and after some further mathematical derivations, we obtain a simpler mathematical expression for the variance-covariance matrix of the sample mean $\overline{\textbf{X}}_t$ as
\begin{align}\label{equ:VCwbar}
 \pmb{\Sigma}_{\overline{\textbf{X}}}=\frac{1}{n} \Bigg[\pmb{\Sigma}_{\textbf{X}} \left(\mathbf{I}_{v}+\Lambda(\pmb{\Phi}^{\intercal})-\frac{1}{n}\Pi(\pmb{\Phi}^{\intercal})\right)+\left(\Lambda(\pmb{\Phi})-\frac{1}{n}\Pi(\pmb{\Phi}) \right) \pmb{\Sigma}^{\intercal}_{\textbf{X}}\Bigg],
\end{align}
where $\pmb{\Phi}$ and $\pmb{\Sigma}_{\textbf{X}}$ are given in \eqref{equ:meanCov} and \eqref{equ:VecSigmaW}, respectively.

\subsection{The VAR($p$) model}

The previous section provides the needed background to obtain the variance-covariance matrix $\pmb{\Sigma}_{\overline{\textbf{X}}}$ for a VAR(1) model as a function of the sample size $n$  and the matrices $\pmb{\Phi}$ and $\pmb{\Sigma}_{\textbf{X}} $ (see equation \eqref{equ:VCwbar}). This section is devoted to deriving a closed-form formula for $\pmb{\Sigma}_{\overline{\textbf{X}}}$ of a VAR($p$) model with any order $p \geq 1$. Let us denote $\mathbf{0}_v$ to be the $v \times v$  zero matrix and $\mathbf{0}$ be a $v(p-1)$-dimensional zero vector. We redefine the VAR($p$) time series $\textbf{X}_t$ of dimension $v$ as the VAR(1) time series $\textbf{Z}_t$ of dimension $vp$ as
\begin{align}\label{equ:expandts}
\mathbf{Z}_{t}=\Big((\textbf{X}_{t}-\pmb{\mu}_{\textbf{X}})^{\intercal},(\textbf{X}_{t-1}-\pmb{\mu}_{\textbf{X}})^{\intercal},\ldots, (\textbf{X}_{t-p+1}-\pmb{\mu}_{\textbf{X}})^{\intercal}\Big)^{\intercal},
\end{align}
where $\textbf{X}_{t}$ is given in \eqref{equ:varp}. By doing this, we have
\begin{align}\label{equ:newreper}
\mathbf{Z}_{t}=\pmb{\Psi} \mathbf{Z}_{t-1}+\boldsymbol{b}_{t},
\end{align}
where $\boldsymbol{b}_{t}=\left(\pmb{\epsilon}_{t}^{\intercal}, \mathbf{0}^{\intercal}\right)^{\intercal}$ is a $vp \times 1$  zero-mean vector with variance-covariance matrix
\begin{align}\label{equ:newcoeff}
\boldsymbol{\Sigma}_{b}=\begin{pmatrix}
\boldsymbol{\Sigma}_{\pmb{\epsilon}}& \mathbf{0}_v& \cdots & \mathbf{0}_v & \mathbf{0}_v\\
\mathbf{0}_v & \mathbf{0}_v & \cdots & \mathbf{0}_v & \mathbf{0}_v \\
\mathbf{0}_v & \mathbf{0}_v & \cdots & \mathbf{0}_v & \mathbf{0}_v \\
\vdots & \vdots & \ddots & \vdots & \vdots \\
\mathbf{0}_v & \mathbf{0}_v & \cdots & \mathbf{0}_v & \mathbf{0}_v
\end{pmatrix}_{vp \times vp},\qquad \text{and} \qquad\pmb{\Psi}=\begin{pmatrix}
\pmb{\Phi}_{1} & \pmb{\Phi}_{2} & \cdots & \pmb{\Phi}_{p-1} & \pmb{\Phi}_{p} \\
\mathbf{I}_v & \mathbf{0}_v & \cdots & \mathbf{0}_v & \mathbf{0}_v \\
\mathbf{0}_v & \mathbf{I}_v & \cdots & \mathbf{0}_v & \mathbf{0}_v \\
\vdots & \vdots & \ddots & \vdots & \vdots \\
\mathbf{0}_v & \mathbf{0}_v & \cdots & \mathbf{I}_v & \mathbf{0}_v
\end{pmatrix}_{vp \times vp}, 
\end{align}
such that $\pmb{\Phi}_{i}$ for $i=1,\ldots,p$ is the $v \times v$ VAR($p$) model's coefficient matrix given in \eqref{equ:varp}. Representing a VAR($p$) with a VAR(1) model with higher dimension helps us to use the derived matrices $\pmb{\Sigma}_{\textbf{X}}$ and $\pmb{\Sigma}_{\overline{\textbf{X}}}$ given in \eqref{equ:VecSigmaW} and \eqref{equ:VCwbar} to calculate their counterparts in the VAR($p$) case. Following a direct approach to derive the variance-covariance matrix of $\mathbf{Z}_{t}$ (by taking the expectation of $\mathbf{Z}_{t}\mathbf{Z}_{t}^{^{\intercal}}$) leads to 
\begin{align}\label{equ:expandtsvarcov}
\pmb{\Sigma}_{\textbf{Z}}=\begin{pmatrix}
\pmb{\Gamma}(0) & \pmb{\Gamma}(1) & \cdots & \pmb{\Gamma}(p-1) \\
\pmb{\Gamma}(-1) & \pmb{\Gamma}(0) & \cdots & \pmb{\Gamma}(p-2) \\
\vdots & \vdots & \ddots & \vdots \\
\pmb{\Gamma}(-p+1)^{\intercal} & \pmb{\Gamma}(-p+2)^{\intercal} & \cdots & \pmb{\Gamma}(0)
\end{pmatrix}_{vp \times vp},
\end{align}
where $\pmb{\Gamma}(k)$ is the  cross-covariance  matrix of the VAR($p$) time series $\textbf{X}_t$ given in \eqref{equ:varp}. Relation \eqref{equ:expandtsvarcov} shows that the   matrices $\pmb{\Gamma}(k)$, and in particular the variance-covariance matrix $\pmb{\Sigma}_{\textbf{X}}=\pmb{\Gamma}(0)$, can be obtained once $\pmb{\Sigma}_{\textbf{Z}}$ is calculated. On the other hand, since $\mathbf{Z}_{t}$ follows a VAR(1) model, from equation \eqref{equ:VecSigmaW} we have  
\begin{align}\label{equ:VecSigmaZ}
\mathrm{Vec}\left(\pmb{\Sigma}_{\textbf{Z}}\right)=\left(\mathbf{I}_{vp \times vp}-\boldsymbol{\Psi} \otimes \boldsymbol{\Psi}\right)^{-1}\mathrm{Vec}\left(\boldsymbol{\Sigma}_{b}\right),
\end{align}
where $\boldsymbol{\Sigma}_{b}$ is given by \eqref{equ:newcoeff}. Accordingly, the variance-covariance matrix $\pmb{\Sigma}_{\textbf{X}}=\pmb{\Gamma}(0)$ of the VAR($p$) time series $\mathbf{X}_{t}$ in \eqref{equ:varp} can be conveniently obtained from relations \eqref{equ:expandtsvarcov} and \eqref{equ:VecSigmaZ}. On the other hand, since $\mathbf{Z}_{t}$ follows a VAR(1) model, the variance-covariance matrix $\pmb{\Sigma}_{\overline{\textbf{Z}}}$ of the sample mean can be obtained from \eqref{equ:VCwbar} based on  $\pmb{\Sigma}_{\textbf{Z}}$ (given in \eqref{equ:VecSigmaZ}) and $\pmb{\Psi}$ (given in \eqref{equ:newcoeff}). It can also be showed that
\begin{align}\label{equ:varcovzbar}
\pmb{\Sigma}_{\overline{\textbf{Z}}}=\begin{pmatrix}
\pmb{\Gamma}_{\overline{\textbf{X}}}(0) & \pmb{\Gamma}_{\overline{\textbf{X}}}(1) & \cdots & \pmb{\Gamma}_{\overline{\textbf{X}}}(p-1) \\
\pmb{\Gamma}_{\overline{\textbf{X}}}(-1) & \pmb{\Gamma}_{\overline{\textbf{X}}}(0) & \cdots & \pmb{\Gamma}_{\overline{\textbf{X}}}(p-2) \\
\vdots & \vdots & & \vdots \\
\pmb{\Gamma}_{\overline{\textbf{X}}}(-p+1) & \pmb{\Gamma}_{\overline{\textbf{X}}}(-p+2) & \cdots & \pmb{\Gamma}_{\overline{\textbf{X}}}(0)
\end{pmatrix}_{vp \times vp},
\end{align} 
so that $\pmb{\Gamma}_{\overline{\textbf{X}}}(0)=\pmb{\Sigma}_{\overline{\textbf{X}}}$ is the variance-covariance matrix of the vector of sample means $\overline{\textbf{X}}$ coming from a VAR($p$) model. Eventually, given the coefficient matrices $\pmb{\Phi_{1}},\ldots,\pmb{\Phi_{p}}$ and the error term variance-covariance matrix $\pmb{\Sigma}_{\varepsilon}$ of the VAR($p$) model in \eqref{equ:varp}, one can obtain $\pmb{\Psi}$ and $\pmb{\Sigma}_{b}$ from \eqref{equ:newcoeff}, and then matrices $\pmb{\Sigma}_{\textbf{X}}$ and $\pmb{\Sigma}_{\overline{\textbf{X}}}$ could be calculated based on equations \eqref{equ:VCwbar} and \eqref{equ:expandtsvarcov}-\eqref{equ:varcovzbar} accordingly. Under the condition $\pmb{\Phi}_1=\ldots=\pmb{\Phi}_p=\textbf{0}$, indicating the absence of any correlation and cross-correlation, we have $\pmb{\Sigma}_{\textbf{X}}=\pmb{\Sigma}_{\varepsilon}$ and $\pmb{\Sigma}_{\overline{\textbf{X}}}=\frac{1}{n}\pmb{\Sigma}_{\textbf{X}}$.

\section{Hotelling's $T^2$ chart}
\label{T2chart}

Hotelling's $T^2$ (or Chi-squared) control chart has been widely implemented in various fields such as manufacturing, finance, and healthcare, where the simultaneous monitoring of the mean vector of multiple correlated variables is crucial for assessing the stability (variability) of processes. Let us consider that in Phase II of the process monitoring, samples of size $n$ are regularly collected from the process output. While the observations within each sample are potentially autocorrelated and/or cross-correlated due to being collected close together in time, the observations in consecutive samples are assumed to be (approximately) independent. This assumption is true in many applications since the time gap between two successive inspections is often large enough to eliminate any dependence between consecutive samples (see for example, Xi and Qiu, 2022). When an assignable cause occurs, the process mean vector shifts from its in-control (IC) value $\pmb{\mu}_{\textbf{X}}^0=(\mu^0_{X_1},\ldots,\mu^0_{X_v})$ to an out-of-control (OOC) value $\pmb{\mu}_{\textbf{X}}^1=(\mu^1_{X_1},\ldots,\mu^1_{X_v})$ where $\mu^1_{X_j}=\mu^0_{X_j}+\delta_j  \sigma_{e_{X_j}}$ for $j=1,\ldots,v$ and $\delta_j$ is a real value. To detect potential shifts in the process mean vector as soon as possible, we propose the $T^2$ statistic with the following form
\begin{align}\label{T2}
T_t^2=(\overline{\textbf{X}}_t-\pmb{\mu}^{0}_{\textbf{X}})^{\intercal} \pmb{\Sigma}_{\overline{\textbf{X}}}^{-1}(\overline{\textbf{X}}_t-\pmb{\mu}^{0}_{\textbf{X}}), \quad t=1,2,\ldots
\end{align}
where the variance-covariance matrix $\pmb{\Sigma}_{\overline{\textbf{X}}}$ for the VAR(1) and VAR$(p)$ models are given in \eqref{equ:VCwbar} and \eqref{equ:varcovzbar}, respectively.  When the process runs in the IC state (i.e., $\overline{\textbf{X}}_t \sim \textbf{MN}_{v}(\pmb{\mu}_{\textbf{X}}^0,\pmb{\Sigma}_{\overline{\textbf{X}}})$), $T^2_t$ for $t=1,2,\ldots$ follows the central chi-square distribution with $v$ degrees of freedom. On the other hand,  when the process mean shifts to $\pmb{\mu}_{\textbf{X}}^1$, $T^2_t$ follows the non-central chi-square distribution with $v$ degrees of freedom and the non-centrality parameter $d=\pmb{\delta}^{\intercal}\pmb{\Sigma}_{\overline{\textbf{X}}}^{-1}\pmb{\delta}$ where $\pmb{\delta}=(\frac{\mu^1_{X_1}-\mu^0_{X_1}}{\sigma_{e_{X_1}}},\ldots,\frac{\mu^1_{X_v}-\mu^0_{X_v}}{\sigma_{e_{X_v}}})$. Using the IC distribution of the $T^2_t$ statistic, the single upper control limit $UCL=\chi^{2}_{v}(\alpha)$ can be applied to check the stability of the process where $\chi^{2}_{v}(\alpha)$ is the upper $\alpha$-th percentile of the chi-square distribution with $v$ degrees of freedom and $\alpha$ is the nominal false alarm rate. The false alarm rate $\alpha$ or equivalently the IC ARL (denoted by $ARL_{0}$) are the most used metrics to statistically design a control chart. In the current problem, we have
\begin{align*}
 \alpha=P(T^2 \geq UCL|\pmb{\mu}_{\textbf{X}}=\pmb{\mu}_{\textbf{X}}^0)=1-F_{\chi^{2}_{v}}(UCL) \quad \text{and} \quad ARL_{0}=\frac{1}{\alpha},
 \end{align*}
where $F_{\chi^{2}_{v}}(\cdot)$ is the cumulative distribution function of the central chi-square distribution with $v$ degrees of freedom. The ARL is the mean of the geometric random variable called Run Length, which is defined as the number of points that are plotted on the chart up to a signal when the process is in the IC or OOC state. The OOC ARL, denoted hereafter by $ARL_{1}$, is the most used performance measure to compare the discrimination power of control charts between the IC and the OOC states. Here, the OOC ARL of the $T^2$ chart can be calculated as
 \begin{align}\label{alr}
 ARL_{1}&=\frac{1}{P(T^2>UCL|\pmb{\mu}_{\textbf{X}}=\pmb{\mu}_{\textbf{X}}^1)}\nonumber\\
 &=\frac{1}{1-F_{\chi^{2}_{(v,d)}}(UCL)},
 \end{align}
 where $F_{\chi^{2}_{(v,d)}}(\cdot)$ is the cumulative distribution function of the non-central chi-square distribution with $v$ degrees of freedom and the non-centrality parameter $d$.

\section{Numerical Study}

To assess the effect of sample size as well as possible autocorrelation and cross-correlation among the measurements on the performance of the $T^2$ chart, this section shows the results of a numerical study based on the $ARL$ metric. First of all, let us set the target IC ARL to $ARL_0=370$, which corresponds to a false alarm rate of $\alpha=0.0027$. During the numerical analysis, we assume that two or three quality characteristics are measured on each item ($v=2$ and $v=3$) and the measurements come from a multivariate normal distribution so that their dependency structure is described by a VAR model of order one ($p=1$). In this setting, we have $UCL=\chi^{2}_{2}(0.0027)=11.827$ for $v=2$ and $UCL=\chi^{2}_{3}(0.0027)=14.154$ for $v=3$. We also consider three different sample sizes $n=3,7,15$. Without loss of generality, we set $\pmb{\mu}^0=\textbf{0}$. Remember that the main aim is to detect possible shifts in $\pmb{\mu}$ as soon as possible. To capture different levels of dependence among measurements, we choose the autocorrelation parameters $\Phi_{X_iX_i}$ for $i=1,\ldots,v$ from the set $\{0.0, 0.3, 0.7\}$ as well as the cross-correlation parameters $\Phi_{X_iX_j}$ $i \neq j,i,j=1,\ldots,v$ from the set $\{0.0, 0.1, 0.3\}$. These choices allow us to evaluate the performance of the chart under varying conditions of weak to strong autocorrelation and cross-correlation along with the case of no autocorrelation and cross-correlation. For the variance-covariance matrix of the error vector $\pmb{\epsilon}_t$, we also consider the forms $\pmb{\Sigma}_{\pmb{\epsilon}}=\begin{pmatrix}1 & \rho \\ \rho&  1\end{pmatrix}$ when $v=2$ and $\pmb{\Sigma}_{\pmb{\epsilon}}=\begin{pmatrix}1 & \rho& \rho\\ \rho&  1 & \rho\\ \rho& \rho& 1 \end{pmatrix} $ when $v=3$ where $\rho \in \{0.0, 0.3, 0.9\}$. This will enable us to explore weak to strong positive linear dependencies among errors which will be translated to the original observations $\textbf{X}_t$ through the VAR model. For example, the choices $\pmb{\Phi}=\begin{pmatrix} 0.0 & 0.0\\0.0&  0.0\end{pmatrix}$ and $\pmb{\Sigma}_{\epsilon}=\begin{pmatrix}1 & 0\\0&  1\end{pmatrix} $ determine two independent sequences of \emph{i.i.d.} variables. To assess the OOC performance of the chart, we consider the shift magnitude of sizes $\delta_i=\delta \in \{0.25,0.50,0.75,1.00,1.50,2.00\}$ for $i=1,\ldots,v$.  \\

\begin{figure}
 \centering
\includegraphics[scale=0.5]{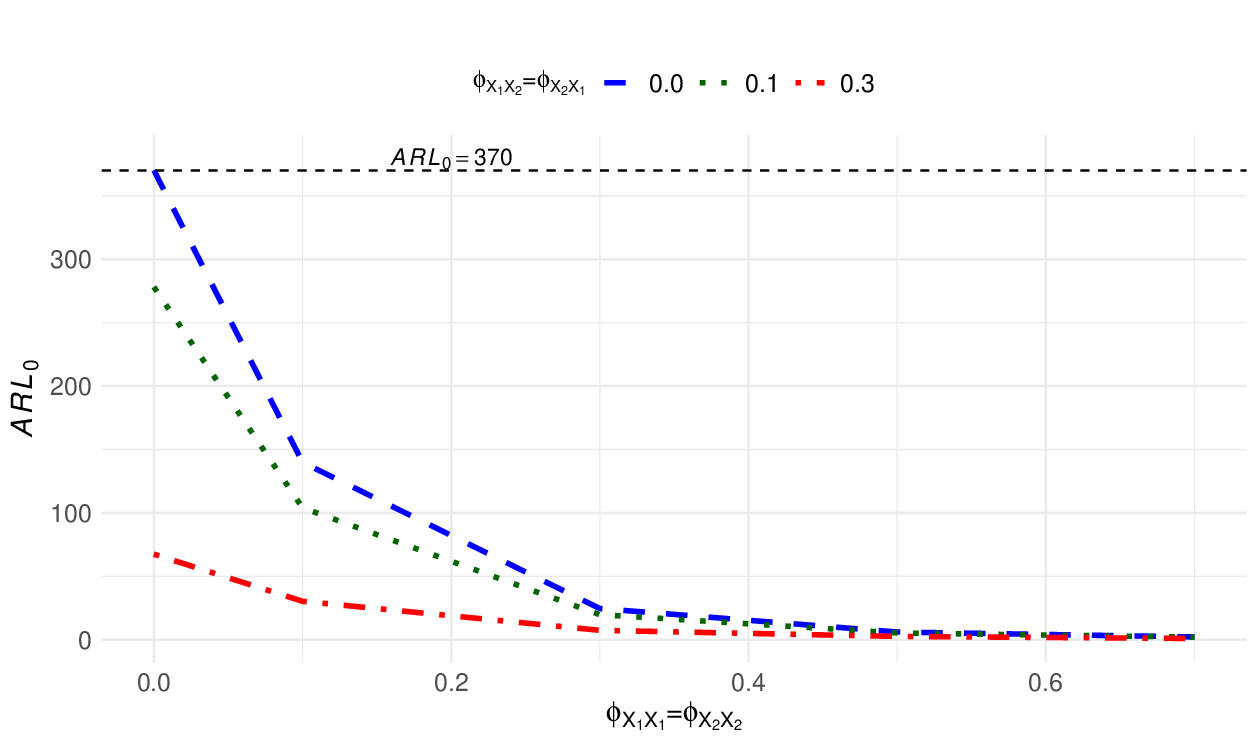}
 \caption{The changes of  ARL$_0$ against the parameters $\Phi_{X_1,X_1}, \Phi_{X_1,X_2}, \Phi_{X_2,X_1},$ and $\Phi_{X_2,X_2}$.}
\label{fig:ARL0}
\end{figure}

\subsection{The effect of autocorrelation on Hotelling (1947)'s $T^2$ chart}
 
We begin by assessing the effect of autocorrelation and cross-correlation among data on the original $T^2$ chart's performance discussed in Hotelling (1947). In particular, we focus on the IC performance ($ARL_0$) of the chart when $v=2$ and $p=1$. More precisely, our goal is to evaluate the IC performance of the $T^2$ chart designed for \emph{i.i.d.} observations when monitoring streams of autocorrelated data. Note that a VAR(1) model with $\pmb{\Phi}=\textbf{0}$ reduces to the model considered in Hotelling (1947). Under this setting, the $T^2_t$ statistic follows $F_{\chi^{2}_{2}}(\cdot)$ distribution and $UCL=\chi^{2}_{2}(0.0027)=11.827$. We generate data from a VAR(1) model with $\Phi_{X_1,X_1}= \Phi_{X_1,X_2}\in \{0.0, 0.1, 0.3, 0.7\}$ and $\Phi_{X_1,X_2}= \Phi_{X_2,X_1}\in \{0.0, 0.1, 0.3\}$. Figure \ref{fig:ARL0} shows the results of this study. It is worth mentioning that Figure \ref{fig:ARL0} does not show the $ARL_1$ values of the chart with respect to the magnitude of a shift in the sense of the traditional ARL plot, rather it displays the $ARL_0$ values of the chart against the autocorrelation and cross-correlation parameters. From this figure, we see that the IC performance of the charts is dramatically affected by the level of autocorrelation and cross-correlation. In the absence of autocorrelation and cross-correlation (i.e., $\Phi_{X_1,X_1}= \Phi_{X_1,X_2}=\Phi_{X_1,X_2}= \Phi_{X_2,X_1}=0$), the IC ARL is 370 which the target value that chart is designed to achieve. However, the $ARL_0$ decreases (the false alarm rate increases) when the level of dependency among measurements gets higher. This means that neglecting the autocorrelation and cross-correlation among observations in the control chart design results in a significantly higher actual false alarm rate than the nominal value upon which the chart is originally based.

\subsection{OOC perfomance evaluation}

\begin{sidewaystable}[]
\centering
\caption{The IC and OOC ARL values when $p=1$ and $v=2$.}
\begin{tabular}{cccccccccccccccccccccc}
\hline\hline
$\pmb{\Sigma}_{\epsilon} \downarrow$ &  & & \multicolumn{3}{c}{$\pmb{\Phi}=\begin{pmatrix}0 & 0\\0&  0\end{pmatrix}$} & & \multicolumn{3}{c}{$\pmb{\Phi}=\begin{pmatrix} 0.0 & 0.1\\0.1&  0.0\end{pmatrix}$}& &\multicolumn{3}{c}{$\pmb{\Phi}=\begin{pmatrix}0.3 & 0.0\\0.0&  0.3\end{pmatrix}$} & &\multicolumn{3}{c}{$\pmb{\Phi}=\begin{pmatrix}0.3 & 0.1\\0.1&  0.3\end{pmatrix}$} & &\multicolumn{3}{c}{$\pmb{\Phi}=\begin{pmatrix}0.7 & 0\\0&  0.7\end{pmatrix}$}  \\
&$\delta \downarrow \ n \rightarrow $& & 3 & 7& 15 & & 3&7&15& & 3&7&15 & & 3&7&15 & & 3&7&15 \\
\hline\multirow{7}{35pt}{$\begin{pmatrix}1 & 0\\0&  1\end{pmatrix} $}
&$0.0$& & 370.0&370.0&370.0&&370.0&370.0&370.0& &370.0 & 370.0& 370.0& &  370.0& 370.0&370.0 & & 370.0& 370.0& 370.0\\
&$0.25$& & 159.6& 77.7& 30.4& & 174.5& 93.3&39.8 & & 208.9&135.3 &70.3 & & 228.1&162.6 & 94.4& &293.5 &267 &221.2 \\
&$0.5$& & 41.1& 11.7&3.4 & &49.3 &15.6 &4.6 & &72.4 & 30.0 & 10.0& & 88.7& 42.7&16.0 & &170.4 & 131.3&82.6 \\
&$0.75$& &12.4 & 3.1&1.3 & &15.5 &4.2 &1.5 & &25.8 & 8.5&2.7 & &34.0&13.0 &4.2 & & 89.5&59.9 & 30.8\\
&$1$& &4.8 &1.5 & 1.0& &6.1 &1.8 & 1.1& & 10.5& 3.3& 1.4& &14.4 &5.0 &1.8 & &46.9 &28.3 & 12.9\\
&$1.5$& & 1.5& 1.0&1.0 & &1.8 &1.0 &1.0 & & 2.8&1.3 &1.0 & &3.8 &1.6 &1.0 & &14.6 & 7.9& 3.4\\
&$2$& &1.0 &1.0 &1.0 & &1.1 &1.0 &1.0 & &1.4 & 1.0& 1.0& & 1.7&1.1 &1.0 & &5.7 &3.1 & 1.6\\
\hline
&$\delta \downarrow \ n \rightarrow $& & 3 & 7& 15 & & 3&7&15& & 3&7&15 & & 3&7&15 & & 3&7&15 \\
\hline\multirow{7}{35pt}{$\begin{pmatrix}1 & 0.3\\0.3&  1\end{pmatrix} $}
&$0.0$& & 370.0&370.0&370.0&&370.0&370.0&370.0& &370.0 & 370.0& 370.0& &  370.0& 370.0&370.0 & & 370.0& 370.0& 370.0\\
&$0.25$& & 187.3& 100.6& 43.3& &201.8 & 118.1&55.3 & & 234.3& 162.7&92.1 & & 251.9& 190.2&119.3 & & 308.7 & 286.2&245.7\\
&$0.5$& &57.1 & 17.8&5.1 & & 67.1& 23.4& 7.1& & 94.6&42.8 &15.3 & &113.0 &59.0 &23.8 & &197.9 &158.6 & 106.1\\
&$0.75$& &18.8 &4.7 & 1.6& &23.3 & 6.4& 2.0& & 37.2&13.0 &4.1 & & 48.0& 19.6&6.5 & &113.9 &79.9 & 43.8\\
&$1$& & 7.4& 2.0&1.1 & &9.4 &2.6 & 1.2& & 16.0&5.1 & 1.8& & 21.7&7.8 &2.6 & &64.3 &40.6 &19.4 \\
&$1.5$& &2.1 &1.1 &1.0 & &2.6 &1.1 & 1.0& &4.3 & 1.6& 1.0& & 5.9& 2.2&1.1 & & 21.9& 12.2&5.2 \\
&$2$& &1.2 &1.0 &1.0 & &1.3 & 1.0& 1.0& &1.9 &1.1&1.0 & &2.4 &1.2 &1.0 & &8.8 & 4.7&2.2 \\
\hline
&$\delta \downarrow \ n \rightarrow $& & 3 & 7& 15 & & 3&7&15& & 3&7&15 & & 3&7&15 & & 3&7&15 \\
\hline\multirow{7}{35pt}{$\begin{pmatrix}1 & 0.9\\0.9&  1\end{pmatrix} $}
&$0.0$& & 370.0&370.0&370.0&&370.0&370.0&370.0& &370.0 & 370.0& 370.0& &  370.0& 370.0&370.0 & & 370.0& 370.0& 370.0\\
&$0.25$& &225.5 &138.1 &68.3 & & 238.7& 157.1&84.1 & &267.0 & 202.4&128.5 & &281.6 & 228.2&158.4 & &326.1 & 309.0&276.5 \\
&$0.5$& & 86.3&31.1 &9.6 & & 98.9& 39.9&13.2 & & 131.3&67.5 & 27.3& & 151.7&88.8 &40.5 & & 235.2& 198.4& 144.2\\
&$0.75$& &32.8 & 8.9& 2.6& &39.7 & 11.9& 3.5& & 59.9&23.4 & 7.6& & 74.6& 34.1& 12.2& &152.6 &114.4 &69.0 \\
&$1$& &13.8 & 3.5& 1.3& &17.3 &4.6 &1.6 & & 28.4&9.5 & 3.0& & 37.2&14.4 &4.7 & &95.4 & 64.7&33.8 \\
&$1.5$& & 3.7& 1.3&1.0 & &4.6 & 1.5&1.0 & &7.9 &2.6 & 1.2& &11.0 &3.8 &1.5 & & 37.7&22.1 &9.8 \\
&$2$& &1.7 & 1.0&1.0 & &2.0 &1.1 &1.0 & &3.1 &1.3 &1.0 & &4.3 & 1.7&1.1 & &16.2 &8.9 & 3.8\\
\hline\hline
\end{tabular}\label{tab1}
\end{sidewaystable}

\begin{sidewaystable}[]
\centering
\caption{The IC and OOC ARL values when $p=1$ and $v=3$.}
\setlength{\tabcolsep}{3pt}
\begin{tabular}{cccccccccccccccccccccc}
\hline\hline
$\pmb{\Sigma}_{\epsilon}$ &  & & \multicolumn{3}{c}{$\pmb{\Phi}=\begin{pmatrix}0 & 0& 0\\0&  0 & 0\\ 0& 0& 0\end{pmatrix}$} & & \multicolumn{3}{c}{$\pmb{\Phi}=\begin{pmatrix}0 & 0.1& 0.1\\ 0.1&  0 & 0.1\\ 0.1& 0.1& 0\end{pmatrix}$}& &\multicolumn{3}{c}{$\pmb{\Phi}=\begin{pmatrix}0.3 & 0& 0\\ 0&  0.3 & 0\\ 0& 0& 0.3\end{pmatrix}$} & &\multicolumn{3}{c}{$\pmb{\Phi}=\begin{pmatrix}0.3 & 0.1& 0.1\\ 0.1&  0.3 & 0.1\\ 0.1& 0.1& 0.3\end{pmatrix}$} & &\multicolumn{3}{c}{$\pmb{\Phi}=\begin{pmatrix}0.7 & 0& 0\\0&  0.7 & 0\\ 0& 0& 0.7\end{pmatrix}$}  \\
&$\delta \downarrow \ n \rightarrow $& & 3 & 7& 15 & & 3&7&15& & 3&7&15 & & 3&7&15 & & 3&7&15 \\
\hline\multirow{7}{65pt}{$\begin{pmatrix}1 & 0& 0\\ 0&  1 & 0\\ 0& 0& 1\end{pmatrix} $}
&$0.0$& & 370.0&370.0&370.0&&370.0&370.0&370.0& &370.0 & 370.0& 370.0& &  370.0& 370.0&370.0 & & 370.0& 370.0& 370.0\\
&$0.25$& & 143.3&63.3&21.9&&175.6&96.2&40.7& &194.3 & 118.8& 56.5& &  236.6& 179.0&110.4 & & 285.2 & 256.4& 207.4\\
&$0.5$& &30.8 &7.7&2.2&&46.8&14.9&4.3& & 58.4&21.6 & 6.6 & &  93.3& 48.8& 18.9& & 154.4 & 114.9&67.8\\
&$0.75$& & 8.2&2.1&1.1&&13.7&3.7&1.4& &18.3 &5.5 &1.8  & & 34.8 & 14.4& 4.8& &  74.3& 47.1&22.3\\
&$1$& &3.1 &1.2&1.0&&5.1&1.6&1.0& &6.9 & 2.2& 1.1 & & 14.2 &5.4& 2.0& & 35.8 & 20.3&8.5\\
&$1.5$& &1.2 &1.0&1.0&&1.5&1.0&1.0& &1.9 &1.1 &1.0 & & 3.5 & 1.6&1.0  & & 9.8 & 5.1&2.2\\
&$2$& & 1.0&1.0&1.0&&1.1&1.0&1.0& &1.1 & 1.0&1.0  & &  1.6&1.1 &1.0 & & 3.7 & 2.1&1.2\\
\hline
&$\delta \downarrow \ n \rightarrow $& & 3 & 7& 15 & & 3&7&15& & 3&7&15 & & 3&7&15 & & 3&7&15 \\
\hline\multirow{7}{65pt}{$\begin{pmatrix}1 & 0.3& 0.3\\ 0.3&  1 & 0.3\\ 0.3& 0.3& 1\end{pmatrix} $}
&$0.0$& & 370.0&370.0&370.0&&370.0&370.0&370.0& &370.0 & 370.0& 370.0& &  370.0& 370.0&370.0 & & 370.0& 370.0& 370.0\\
&$0.25$& &194.0 &104.3&43.7&&224.5&144.2&73.2& & 241.1&169.0 &95.3 & & 276.3 &227.6 & 160.0& & 313.1 & 291.6& 252.3\\
&$0.5$& & 58.2&17.1&4.7&&82.0&31.2&9.6& & 97.9&43.1 &  14.7& & 140.9 &84.7 &38.5 & &204.8  &164.8 &110.1\\
&$0.75$& &18.2 &4.3&1.5&&28.9&8.3&2.5& & 37.3&12.3 & 3.7 & & 64.4 &30.3 &10.7 & &  118.2& 82.5&44.3\\
&$1$& &6.8 &1.8&1.0&&11.4&3.1&1.3& & 15.4&4.6 &1.6  & & 29.8 &12.1 &4.0 & & 65.9 & 40.9&18.8\\
&$1.5$& &1.9 &1.0&1.0&&2.9&1.2&1.0& &3.8 & 1.4& 1.0& & 7.9 & 3.0&1.3 & &21.4  & 11.6&4.7\\
&$2$& &1.0 &1.0&1.0&&1.4&1.0&1.0& & 1.7& 1.0&1.0  & & 3.0 & 1.4&1.0 & &  8.2&4.3 &1.9\\
\hline
&$\delta \downarrow \ n \rightarrow $& & 3 & 7& 15 & & 3&7&15& & 3&7&15 & & 3&7&15 & & 3&7&15 \\
\hline\multirow{7}{65pt}{$\begin{pmatrix}1 & 0.9& 0.9\\ 0.9&  1 & 0.9\\ 0.9& 0.9& 1\end{pmatrix} $}
&$0.0$& & 370.0&370.0&370.0&&370.0&370.0&370.0& &370.0 & 370.0& 370.0& &  370.0& 370.0&370.0 & & 370.0& 370.0& 370.0\\
&$0.25$& & 249.0&163.1&85.4&&273.5&204.2&126.2& &286.1 & 227.3&152.9& & 311.3 & 275.9&219.1& &  335.6& 321.7&294.4 \\
&$0.5$& & 106.4&40.0&12.2&&137.1&65.5&24.1& &155.9 & 84.5& 35.1 & & 201.1 & 140.4&77.3 & & 257.8 & 223.5&169.4\\
&$0.75$& &42.1 &11.2&3.1&&61.6&21.2&6.2& & 75.4&30.2 & 9.6 & & 114.7 &64.0 &26.6 & & 178.1 &137.7 &86.2\\
&$1$& & 17.7&4.2&1.4&&28.3&8.1&2.4& & 36.5&12.0 &3.6  & & 63.2 &29.6 &10.4 & &116.6  &81.1 &43.3\\
&$1.5$& & 4.4&1.4&1.0&&7.4&2.2&1.1& &10.0 & 3.0& 1.3& &  20.2& 7.8&2.7  & &  48.2& 28.5&12.4\\
&$2$& &1.9 &1.0&1.0&&2.8&1.2&1.0& & 3.7&1.4 &1.0  & & 7.7 &3.0 & 1.3& & 20.9 & 11.2&4.6\\
\hline\hline
\end{tabular}\label{tab2}
\end{sidewaystable}

In this section, we assess the performance of the $T^2$ chart discussed in Section \ref{T2chart}  in terms of the ability to detect the shifts in the process mean vector $\pmb{\mu}$ based on the $ARL_1$ metric. The results of this study when $v=2$ and $v=3$ are presented in Tables  \ref{tab1} and \ref{tab2}, respectively. In particular, the detection power of the monitoring techniques is investigated in terms of the chart and process parameters. The insights from these results help us understand how various factors such as shift size, sample size, level of autocorrelation and cross-correlation, and data dimensionality impact the performance of control charts in detecting OOC conditions. According to these tables, we can see that the control chart is ARL-unbiased, i.e., $ARL_0 > ARL_1$, for the shifts investigated in this study. It can also be observed that as the shift size $\delta$ increases, $ARL_1$ decreases across all sample sizes and model parameters. For larger shifts ($\delta \geq 1.5$), the ARL values quickly converge to 1, indicating that the control chart detects the shift almost immediately. The second noticeable result is that the charts detect shifts in $\pmb{\mu}$ sooner when $n$ increases. This is expected because a larger sample size provides more information about the process and consequently enhances the sensitivity of the chart, leading to faster detection of OOC conditions. Table \ref{tab1} and \ref{tab2} show that the detection ability of the chart deteriorates when the degree of association among data increases. This is equivalent to say that the statistical performance deteriorates with the degree of dependency. This suggests that the presence of high correlation and/or autocorrelation makes it more challenging to detect shifts, requiring more observations before a signal is given. 

%Finally, comparing the tables for $v = 2$ and $v = 3$, it is observed that higher dimensions generally lead to higher $ARL_1$ values for the same $\delta, n, \pmb{\Phi},$ and $\pmb{\Sigma}_{\pmb{\epsilon}}$. The increase in $ARL_1$ values with dimensionality indicates that as the complexity of the data increases, the control chart's sensitivity to shifts decreases.

\subsection{Comparison study}

This subsection is devoted to comparing the performance of Hotteling $T^2$ chart discussed in Section 3 to an alternative chart. To do this, we first introduce the alternative monitoring technique for a VAR process and then performance comparisons will be provided, accordingly. As mentioned earlier, Jarrett and Pan (2007) and Pan and Jarrett (2012) proposed to monitor residuals of a VAR($p$) model through Hotelling $T^2$ chart. Their proposed method is applicable to individual observations ($n=1$) and we easily adapt it here for the comparison purpose to the case $n  \geq 1$. Since all the process parameters are assumed to be \emph{known}  from a Phase I analysis, the predicted values of the time series are calculated based on the known parameters and the residuals can be calculated accordingly (Vanhatalo and Kulahci (2015)).  Thus, the residuals can be mathematically calculated as $\pmb{e}_t=\textbf{X}_t-\tilde{\textbf{X}}_{t}=\textbf{X}_t-\textbf{C}-\pmb{\Phi}\textbf{X}_{t-1}$. Not that the variance-covariance matrix of the residuals is the same as that of the error term vector, i.e., $\pmb{\Sigma}_{\pmb{\epsilon}}$. Finally, a $T^2$ chart with monitor statistic
\begin{align}\label{T2E}
T_t^2=\overline{\pmb{e}}_t^{\intercal} \pmb{\Sigma}_{\pmb{\bar{e}}}^{-1}\overline{\pmb{e}}_t
\end{align}
can be applied to detect the possible shifts in the mean vector $\pmb{\mu}$. In relation \eqref{T2E}, $\overline{\pmb{e}}_t$  is the vector of the averages of residuals at time $t$ and $\pmb{\Sigma}_{\pmb{\bar{e}}}=\frac{1}{n}\pmb{\Sigma}_{\pmb{\epsilon}}$. It is important to mention that since we assume a perfect model fit, this approach is expected to provide the best-case scenario for the residual-based $T^2$ chart.

\begin{figure}[]
\centering
 \subfigure[Case I.]{%
 \resizebox*{6.5cm}{!}{\includegraphics{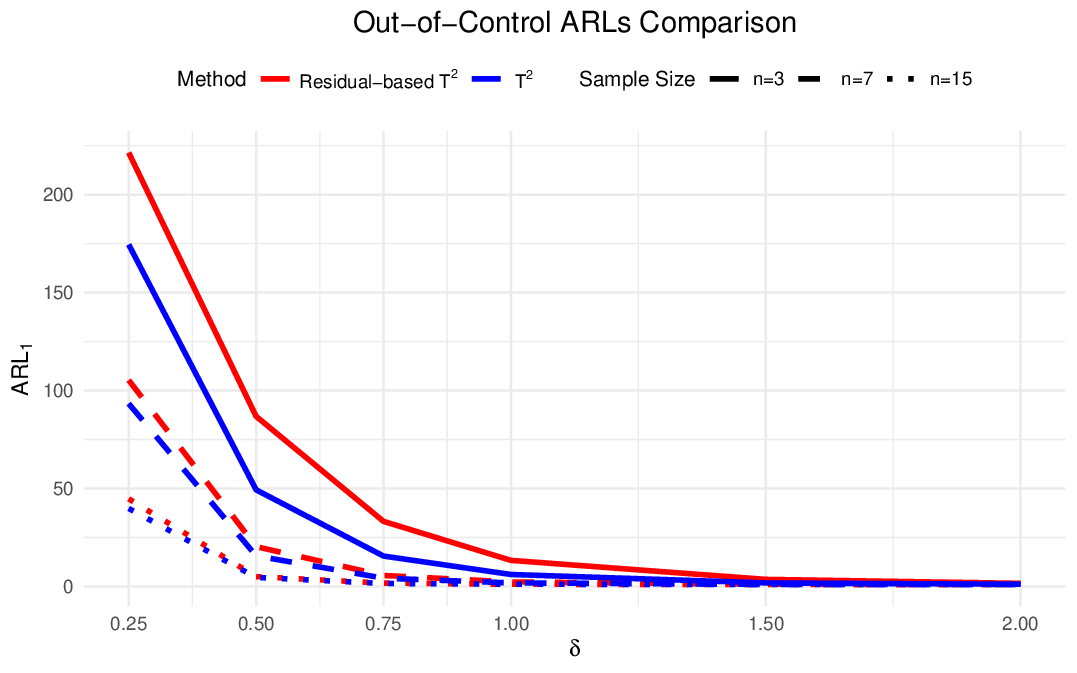}}}
 \subfigure[Case II. ]{%
 \resizebox*{6.5cm}{!}{\includegraphics{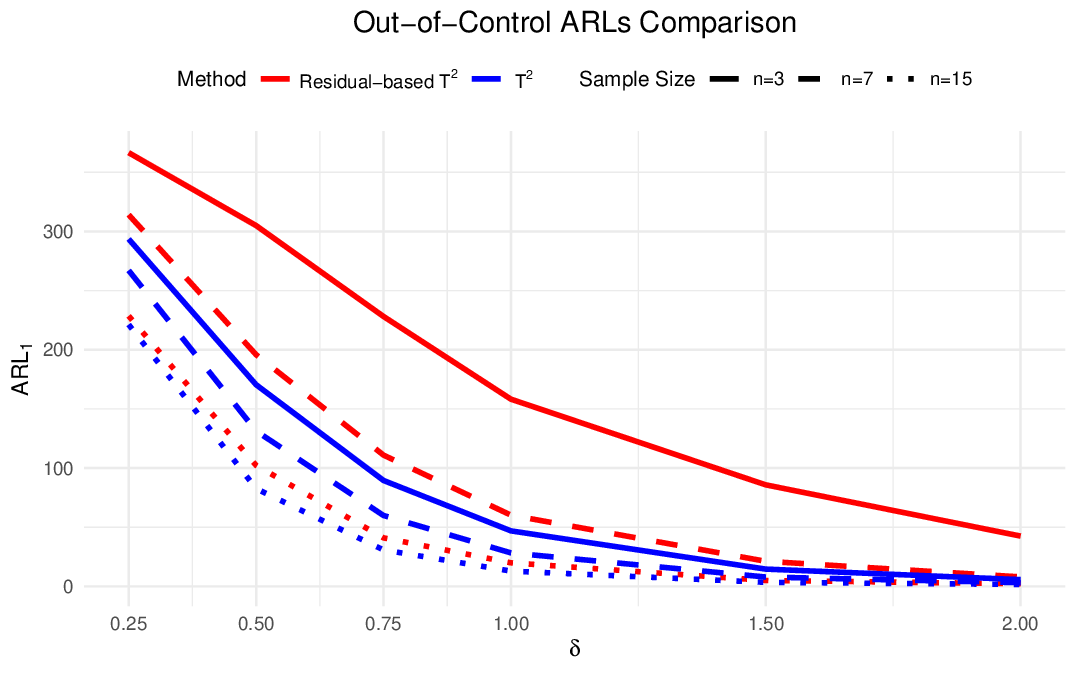}}}
 \subfigure[Case III. ]{%
 \resizebox*{6.5cm}{!}{\includegraphics{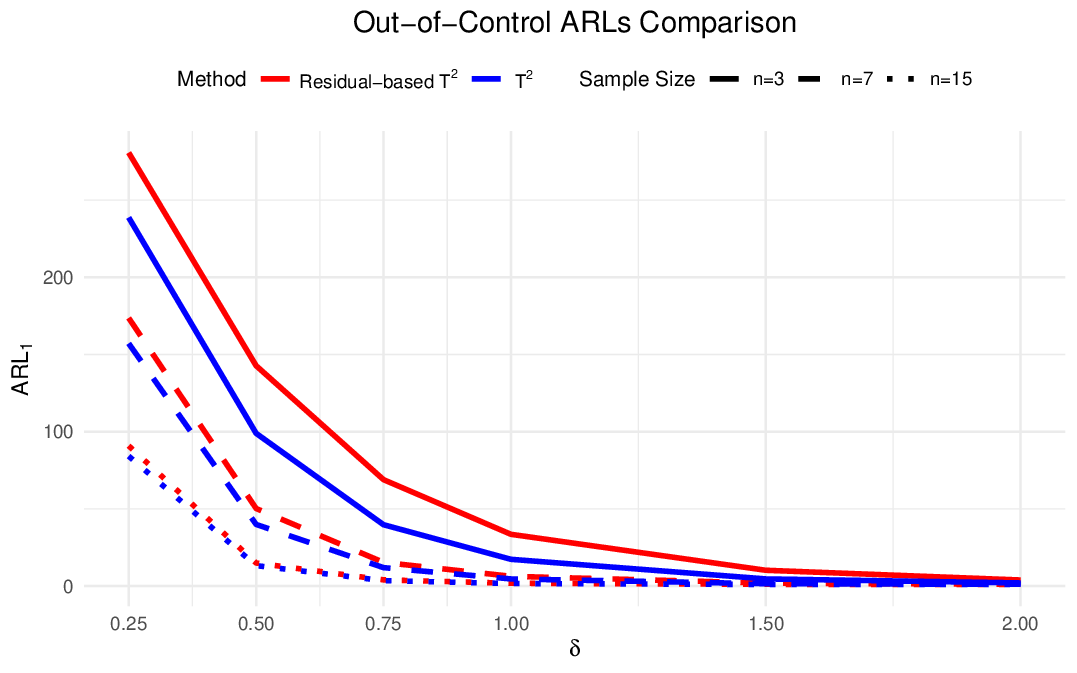}}}
 \subfigure[Case IV.]{%
 \resizebox*{6.5cm}{!}{\includegraphics{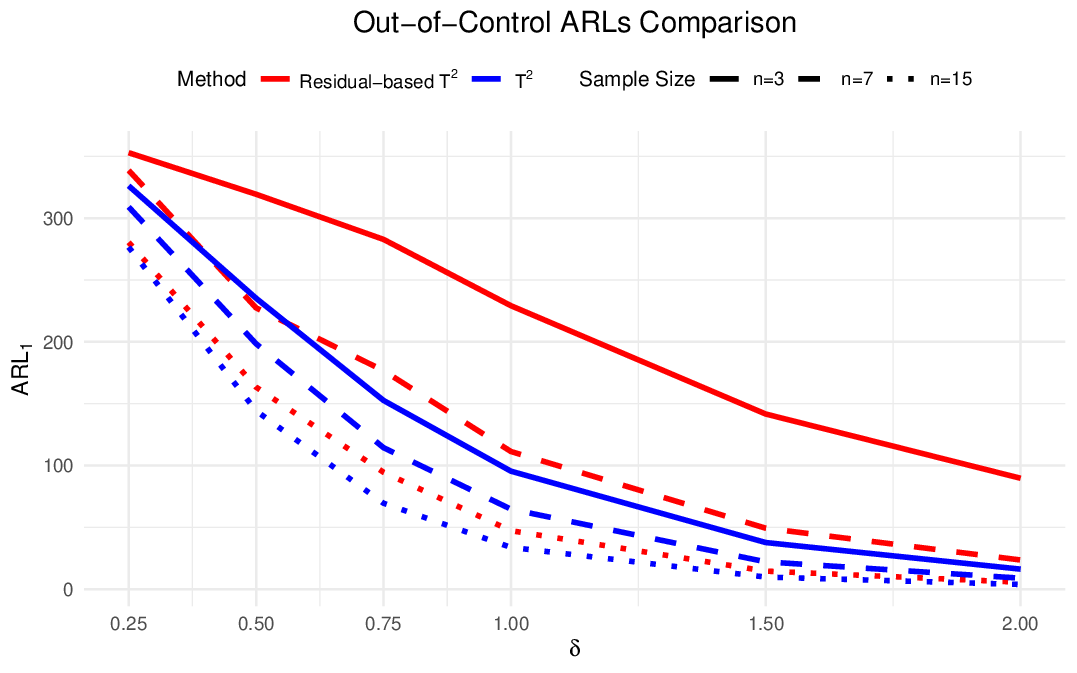}}}
  \subfigure[Case V.]{%
 \resizebox*{6.5cm}{!}{\includegraphics{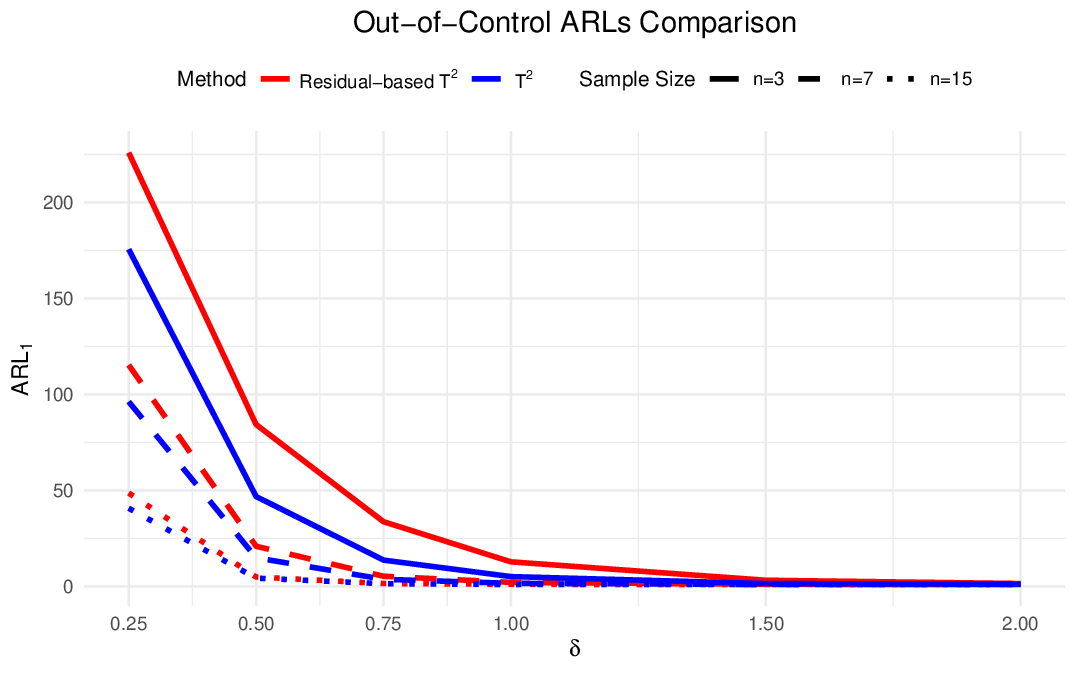}}}
 \subfigure[Case VI.]{%
 \resizebox*{6.5cm}{!}{\includegraphics{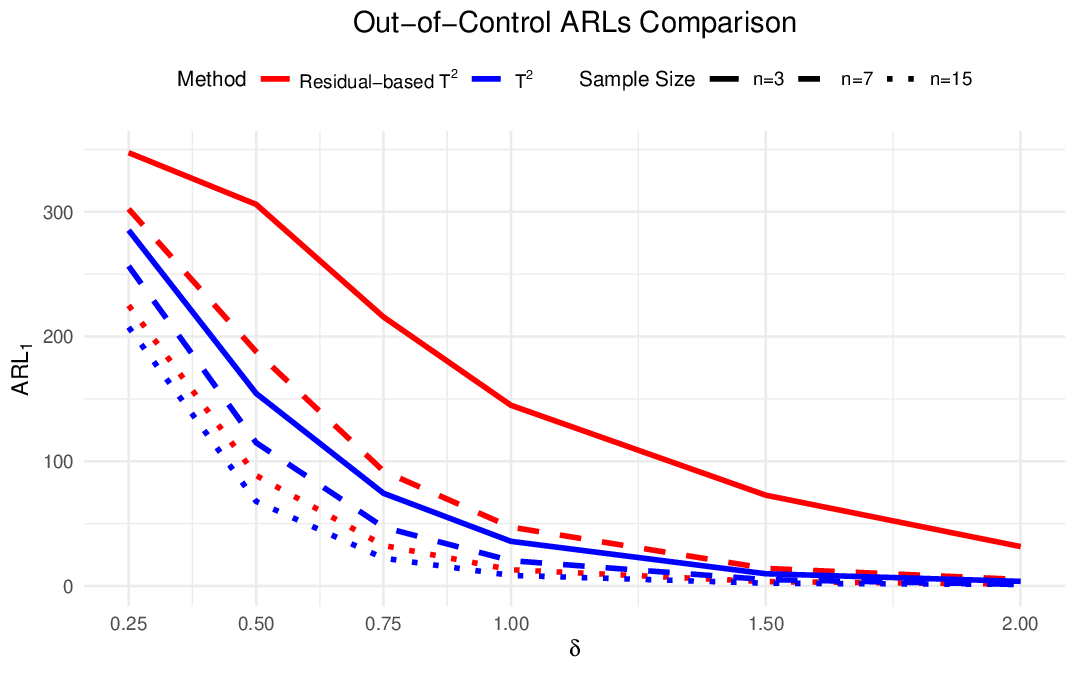}}}
 \subfigure[Case VII.]{%
 \resizebox*{6.5cm}{!}{\includegraphics{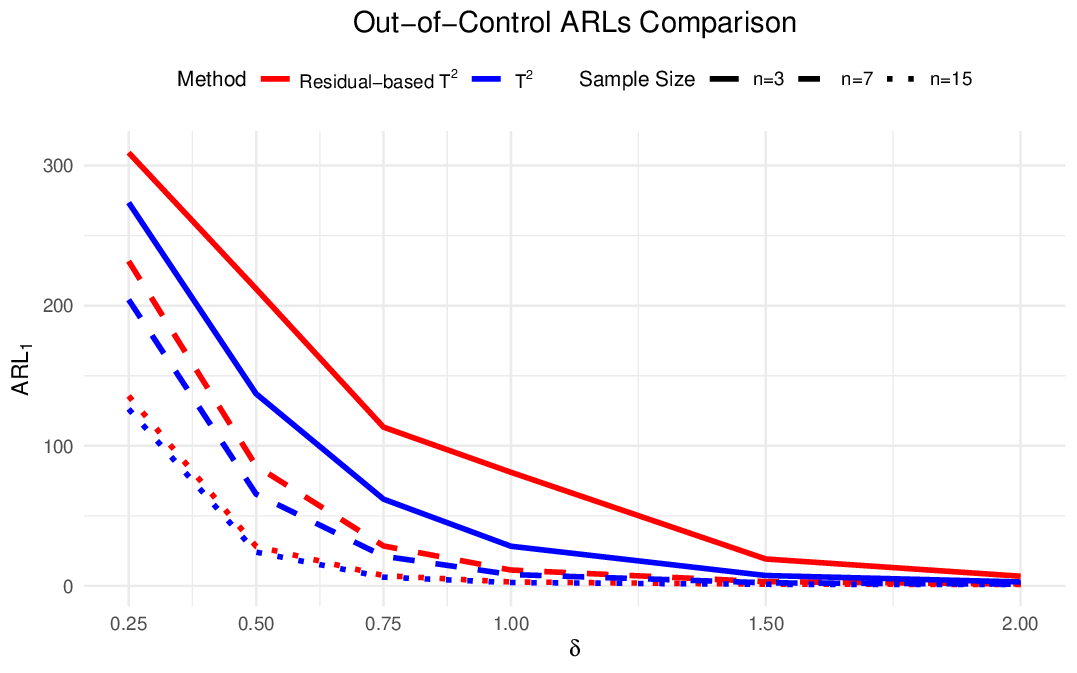}}}
 \subfigure[Case VIII.]{%
 \resizebox*{6.5cm}{!}{\includegraphics{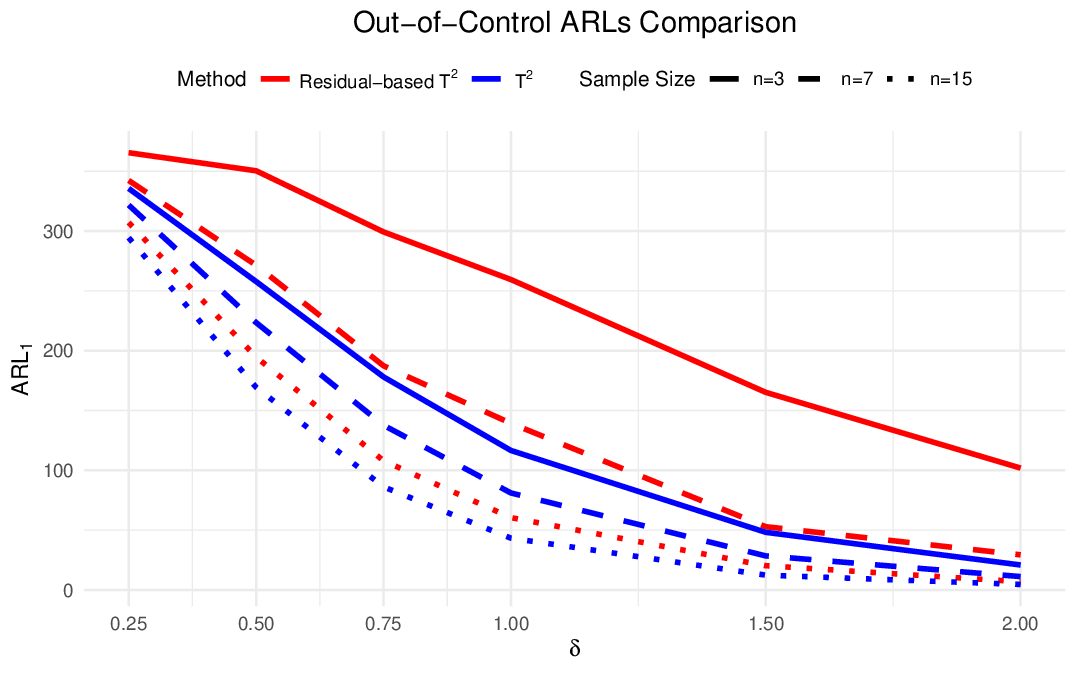}}}
 \caption{The $ARL_1$ values of the $T^2$ chart based on the original observations (blue) against the residual-based $T^2$ chart (red).}
\label{fig:ARL1}
\end{figure}

The OOC ARL is the most used performance measure to compare the discrimination power of control charts between the IC and the OOC states. By examining the $ARL_1$ values in these cases, we can gain insights into the performance and effectiveness of the control charts under various scenarios and deviations from the IC condition. The $ARL_1$ can be calculated from \eqref{alr}. Given a shift of size $\pmb{\delta}$, between two charts with the same $ARL_0$, the chart with the lower $ARL_1$ is preferable because it detects OOC condition more quickly on average. To compare the two charts  for different sample sizes $n=3,7,15$, we consider the following parameter settings: Case I: $v=2, \pmb{\Phi}=\begin{pmatrix} 0.0 & 0.1\\0.1&  0.0\end{pmatrix}, \pmb{\Sigma}_{\pmb{\epsilon}}=\begin{pmatrix}1 & 0\\0&  1\end{pmatrix} $, Case II: $ \pmb{\Phi}=\begin{pmatrix} 0.7 & 0.0\\0.0&  0.7\end{pmatrix}, \pmb{\Sigma}_{\pmb{\epsilon}}=\begin{pmatrix}1 & 0\\0&  1\end{pmatrix} $, Case III: $\pmb{\Phi}=\begin{pmatrix} 0.0 & 0.1\\0.1&  0.0\end{pmatrix}, \pmb{\Sigma}_{\pmb{\epsilon}}=\begin{pmatrix}1 & 0.9\\0.9&  1\end{pmatrix} $, Case IV: $\pmb{\Phi}=\begin{pmatrix} 0.7 & 0.0\\0.0&  0.7\end{pmatrix}, \pmb{\Sigma}_{\pmb{\epsilon}}=\begin{pmatrix}1 & 0.9\\0.9&  1\end{pmatrix} $ when $v=2$ and  Case V: $\pmb{\Phi}=\begin{pmatrix}0 & 0.1& 0.1\\ 0.1&  0 & 0.1\\ 0.1& 0.1& 0\end{pmatrix}, \pmb{\Sigma}_{\pmb{\epsilon}}=\begin{pmatrix}1 & 0& 0\\ 0&  1 & 0\\ 0& 0& 1\end{pmatrix}$,  Case VI: $\pmb{\Phi}=\begin{pmatrix}0.7 & 0.0& 0.0\\ 0.0&  0.7 & 0.0\\ 0.0& 0.0& 0.7\end{pmatrix}, \pmb{\Sigma}_{\pmb{\epsilon}}=\begin{pmatrix}1 & 0& 0\\ 0&  1 & 0\\ 0& 0& 1\end{pmatrix}$, Case VII: $\pmb{\Phi}=\begin{pmatrix}0 & 0.1& 0.1\\ 0.1&  0 & 0.1\\ 0.1& 0.1& 0\end{pmatrix}, \pmb{\Sigma}_{\pmb{\epsilon}}=\begin{pmatrix}1 & 0.9& 0.9\\ 0.9&  1 & 0.9\\ 0.9& 0.9& 1\end{pmatrix}$, and the Case VIII: $\pmb{\Phi}=\begin{pmatrix}0.7 & 0.0& 0.0\\ 0.0&  0.7 & 0.0\\ 0.0& 0.0& 0.7\end{pmatrix}, \pmb{\Sigma}_{\pmb{\epsilon}}=\begin{pmatrix}1 & 0.9& 0.9\\ 0.9&  1 & 0.9\\ 0.9& 0.9& 1\end{pmatrix}$ when $v=3$. These settings have been chosen to compare the performance of both charts in the presence of different sample sizes and process dimensions when the correlation and/or autocorrelation level increases. The results of this study are presented in Figure \ref{fig:ARL1}. According to this figure, the $T^2$ chart that uses the original observations uniformly performs better than the residual-based $T^2$ chart across all cases showing faster detection of OOC conditions. In particular, for small shifts ($\delta \leq 1.0$), our chart demonstrates significantly lower $ARL_1$ compared to the residual-based chart. For example, in Case IV with a shift of size $\delta=1.0$, the OOC ARL of our chart and the residual-based chart are, respectively, $95.4$ and $227.9$  when $n=3$, $64.7$ and $121.2$ when $n=7$, and $33.8$ and $49.4$ when $n=15$.  Generally, as the shift size increases, both charts' ARLs converge to one, but our chart remains faster in signalling. Though these figures demonstrate the effect of autocorrelation on detection ability, showing that higher autocorrelation generally increases ARLs, the increase is more pronounced in the residual-based chart. This indicates that the performance of our chart is less impacted (more robust) by the association among data. Another important trend observable in the tables is that the difference in performance between the two approaches is more influenced by changes in $\pmb{\Phi}$ than by changes in $\pmb{\Sigma}_{\epsilon}$. This is likely because the residual-based approach incorporates  $\pmb{\Sigma}_{\epsilon}$ directly into its construction, thereby accounting for the covariance structure of the errors. However, it does not similarly account for $\pmb{\Phi}$ which governs the autocorrelation and cross-correlation in the data. Therefore, changes in $\pmb{\Phi}$ have a more pronounced impact on the relative performance of the residual-based approach compared to changes in $\pmb{\Sigma}_{\epsilon}$. Furthermore, it can be observed that for both cases $v=2$ and $v=3$, our chart maintains lower OOC ARLs. This indicates that the $T^2$ chart that uses original observations is more effective in higher dimensions, maintaining quick detection capabilities even as the complexity of the data increases. In summary, this ARL analysis suggests that the $T^2$ chart based on original observations is generally more effective and reliable than the residual-based approach for quick detection of OOC conditions across a wide range of scenarios. This poor performance of the residual-based control chart can be attributed to the loss of information or the dependence left among residuals (Yao et al. (2023)).

\begin{figure}[]
\centering
 \subfigure[Case I.]{%
 \resizebox*{7.9cm}{!}{\includegraphics{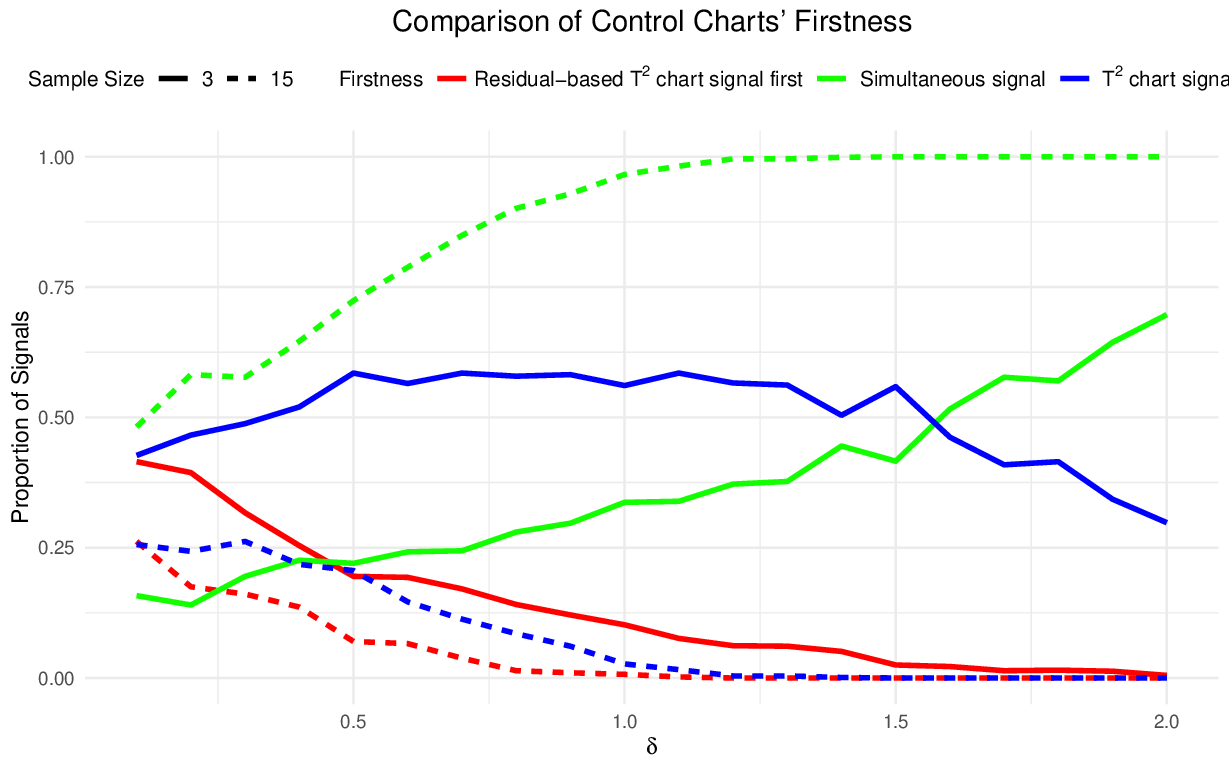}}}
 \subfigure[Case II. ]{%
 \resizebox*{7.9cm}{!}{\includegraphics{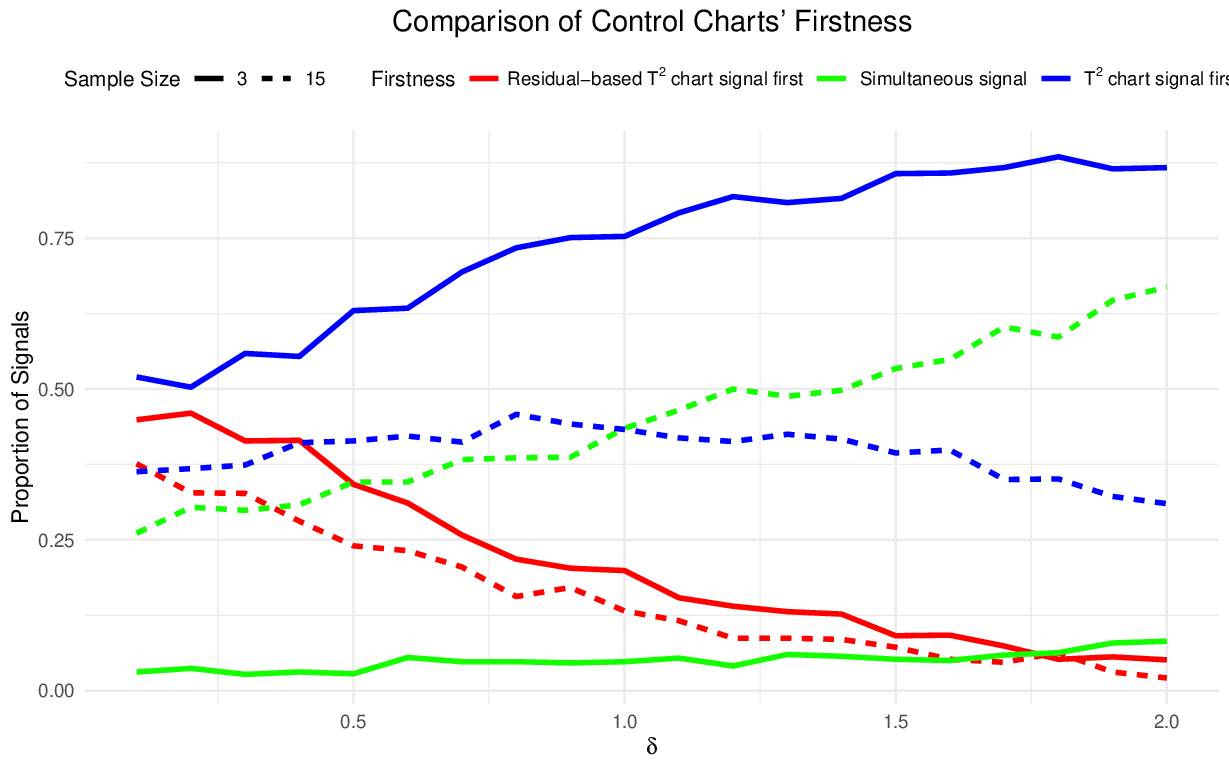}}}
 \subfigure[Case III. ]{%
 \resizebox*{7.9cm}{!}{\includegraphics{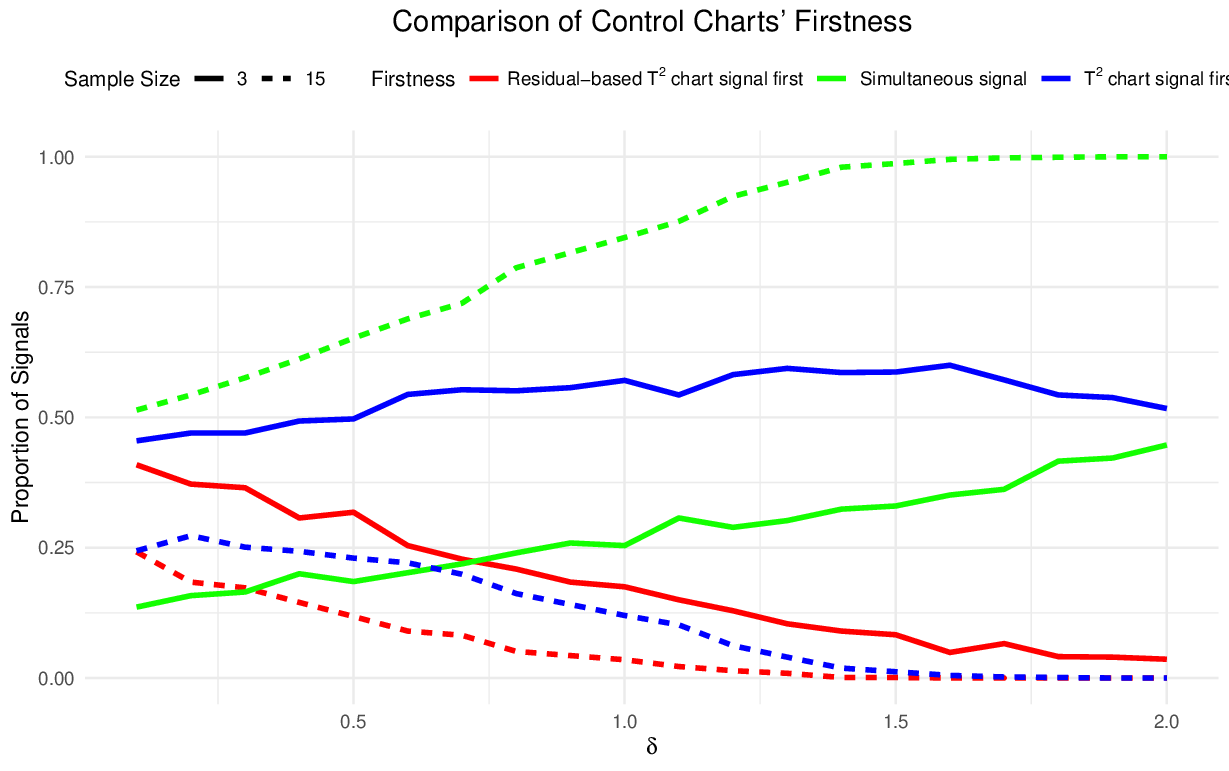}}}
 \subfigure[Case IV.]{%
 \resizebox*{7.9cm}{!}{\includegraphics{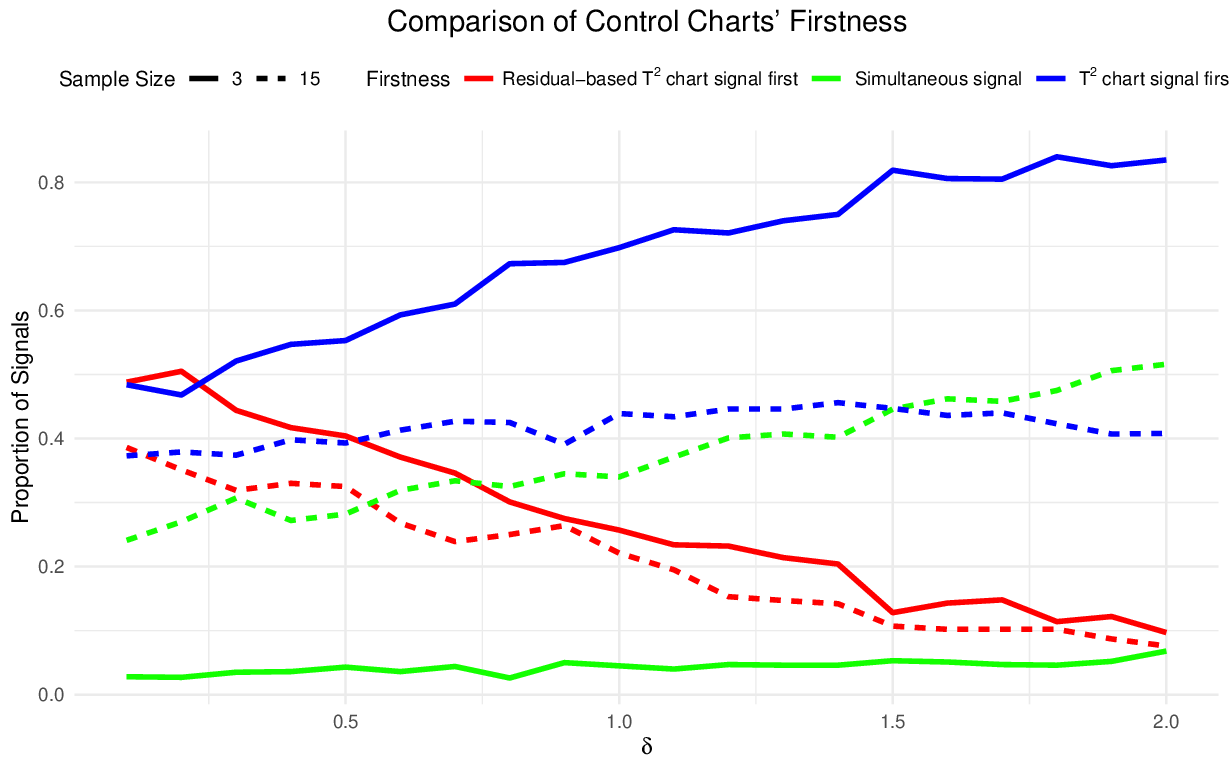}}}
 \caption{ Head-to-head comparison of the $T^2$ chart based on the original observations (blue) chart and the residual-based $T^2$ chart using $p_1$ (blue), $p_2$ (red), and $p_3$ (green) metrics.}
\label{fig:fts}
\end{figure}

When the OOC ARL of a chart is smaller than an alternative chart, it means that on average the first chart will signal before the alternative one. However, it does not answer the question "How likely it is that the first chart will signal before the alternative chart". Recently, Rigdon et al. (2024) discussed this issue and introduced an alternative performance metric called the "first-to-signal" or "firstness" criterion for control chart comparison. To adopt their criterion in our problem, let us define $R_1$ and $R_2$ as the first-time index for which the $T^2$ chart based on the original observations and the residual-based $T^2$ chart, respectively, raise an OOC signal using the same stream of data. It can be seen that $R_1$ and $R_2$ are discrete and dependent random variables.  Having this definition for a given sample, there are three possibilities:  our chart signals before the residual-based chart ($R_1 < R_2$), the residual-based chart signals before our chart  ($R_2 < R_1$), and both methods signals at the same time ($R_1=R_2$). Given this, Rigdon et al. (2024) defined the following probabilities
 \begin{align}\label{probs}
p_1=P(R_1 < R_2), \quad p_2=P(R_2 < R_1), \quad \text{and} \quad p_3=P(R_1=R_2), \quad \text{so that} \quad p_1+p_2+p_3=1. 
 \end{align}
 
 According to the first-to-signal criterion, in a head-to-head comparison, our chart beats the residual-based $T^2$ chart when $p_1 > p_2$.  In this study, we also follow this approach to compare both charts with the same IC ARL equal to 370. Not that the probabilities in \eqref{probs} are estimated through simulations of 10,000 replications each. We did this comparison study for Cases I-IV ($v=2$) of parameters settings introduced in the ARL analysis when $n=3,15$. The results are presented in Figure \ref{fig:fts}. In this figure, $p_1$ is in blue, $p_2$ is in red, and $p_3$ is in green. The charts are evaluated under different parameter settings (Cases I-IV) with varying sample sizes ($n=3,15$) and shift sizes ($\delta=0.1(0.1)2.0$). This figure shows that in all scenarios, $p_1$ (blue) is consistently higher than $p_2$ (red), indicating that the $T^2$ chart based on the original measurements signals before the residual-based $T^2$ chart more frequently. For both $n=3$ and $n=15$, the trends are consistent, with the original $T^2$ chart generally outperforming the residual-based chart in terms of signalling first. For example, in Case IV with a shift of size $\delta=1.0$ and sample size $n=3$, we see that the $T^2$ chart with original observations wins (in the sense that it signals first) 67.7 $\%$ of the time, the residual-based $T^2$ chart signals first 27.4 $\%$ of the time and 4.9 $\%$ of the time they signal simultaneously. In addition, as $\delta$ increases, $p_1$ tends to increase, suggesting that the original $T^2$ chart becomes more effective at detecting larger shifts earlier than the residual-based chart. Figure 3 visually demonstrates the effectiveness of the original $T^2$  chart over the residual-based chart across various conditions. The original $T^2$ chart consistently signals OOC conditions faster, highlighting its robustness and reliability in process monitoring. The metrics $p_1, p_2,$ and $p_3$ provide a comprehensive comparison that shows the superiority of the original $T^2$ chart in the context of the first-to-signal criterion.

%For larger shifts, the Shewhart chart is more likely to signal first. The bottom panel in Figure 1 shows that the EWMA chart is uniformly more likely to signal before the CUSUM chart, although, for shifts beyond d ¼ 1, the most likely scenario is that both charts signal simultaneously. This may be due to the use of exact limits for the EWMA chart.

\section{Illustrative examples}
\label{sec:illustrative} 

This section demonstrates some applications and illustrates the step-by-step implementation procedures of Hotelling's $T^2$   control charts with VAR($p$) observations in two real-world processes. To this end, we consider two examples, a steel sheet rolling process and a chemical process.

\subsection{Steel sheet  rolling process: VAR(1) observations}

\subsubsection*{Process description}

The production of metallic sheets is an important step in countless manufacturing processes in the automotive, aerospace, oil and gas industries, etc. In this regard, the cold rolling process has been extensively used for steel sheet fabrication: monitoring the thickness of the sheets during the process is a very important quality control activity in these forming processes. In fact, the mechanical properties of cold-rolled sheets, such as their strength, are highly affected by the thickness stability within the product (Younes et al. (2006)). For this reason, thickness is a critical quality output of the rolling processes (Sparks et al. (1997) and Sparks (2015)). However, several manufacturing operating conditions can lead to a shift in a sheet's thickness, such as the reduction ratio, rolling speed, roll condition, strip temperature, and raw material composition (Ubici et al. (2001)). Herein, developing effective monitoring techniques to detect the occurrence of an assignable cause as soon as possible is of significant value in quality monitoring and improvement of these processes.

\begin{figure}
\centering
 \subfigure[A steel sheet rolling process.]{%
 \resizebox*{7.5cm}{!}{\includegraphics{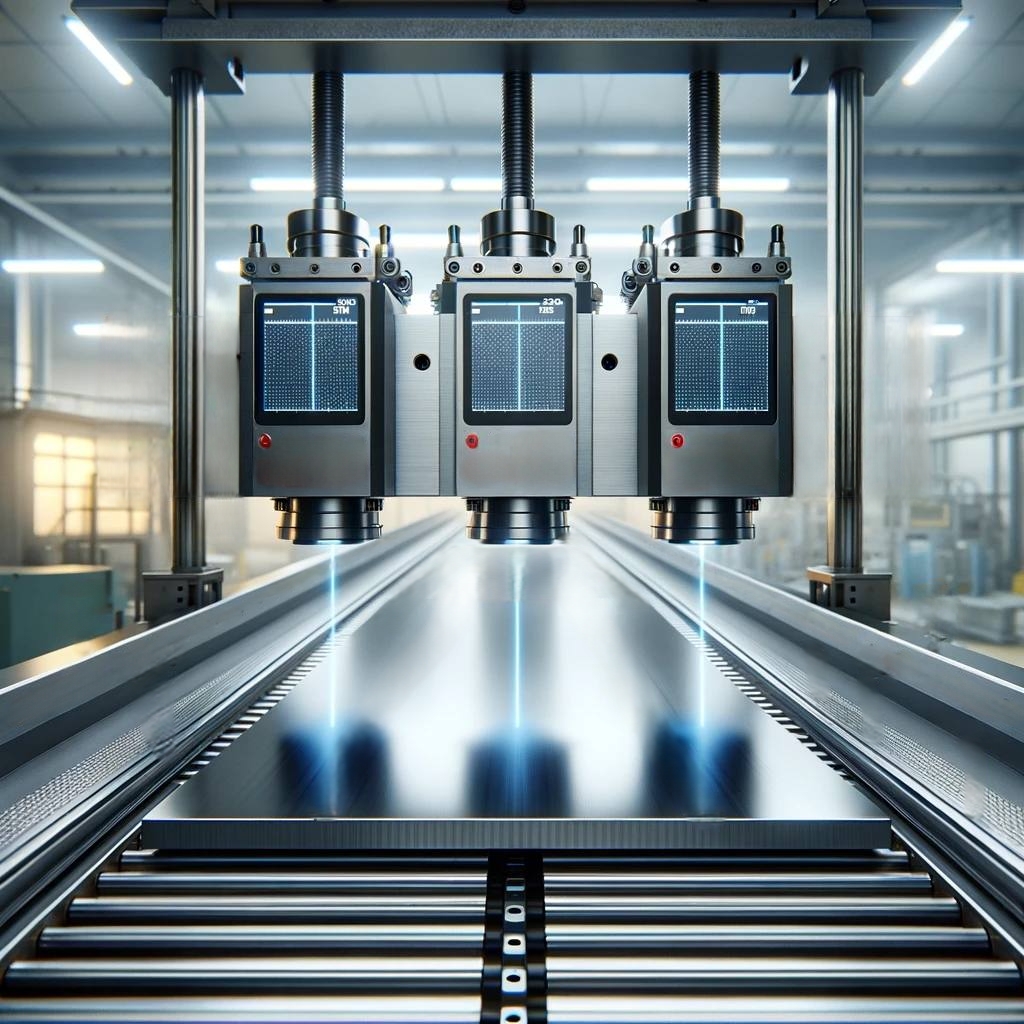}}}
 \subfigure[The location of thickness measurements on the steel sheet   and correlation, autocorrelation, and cross-correlation  between  neighbour spots.]{%
 \resizebox*{7.5cm}{!}{\includegraphics{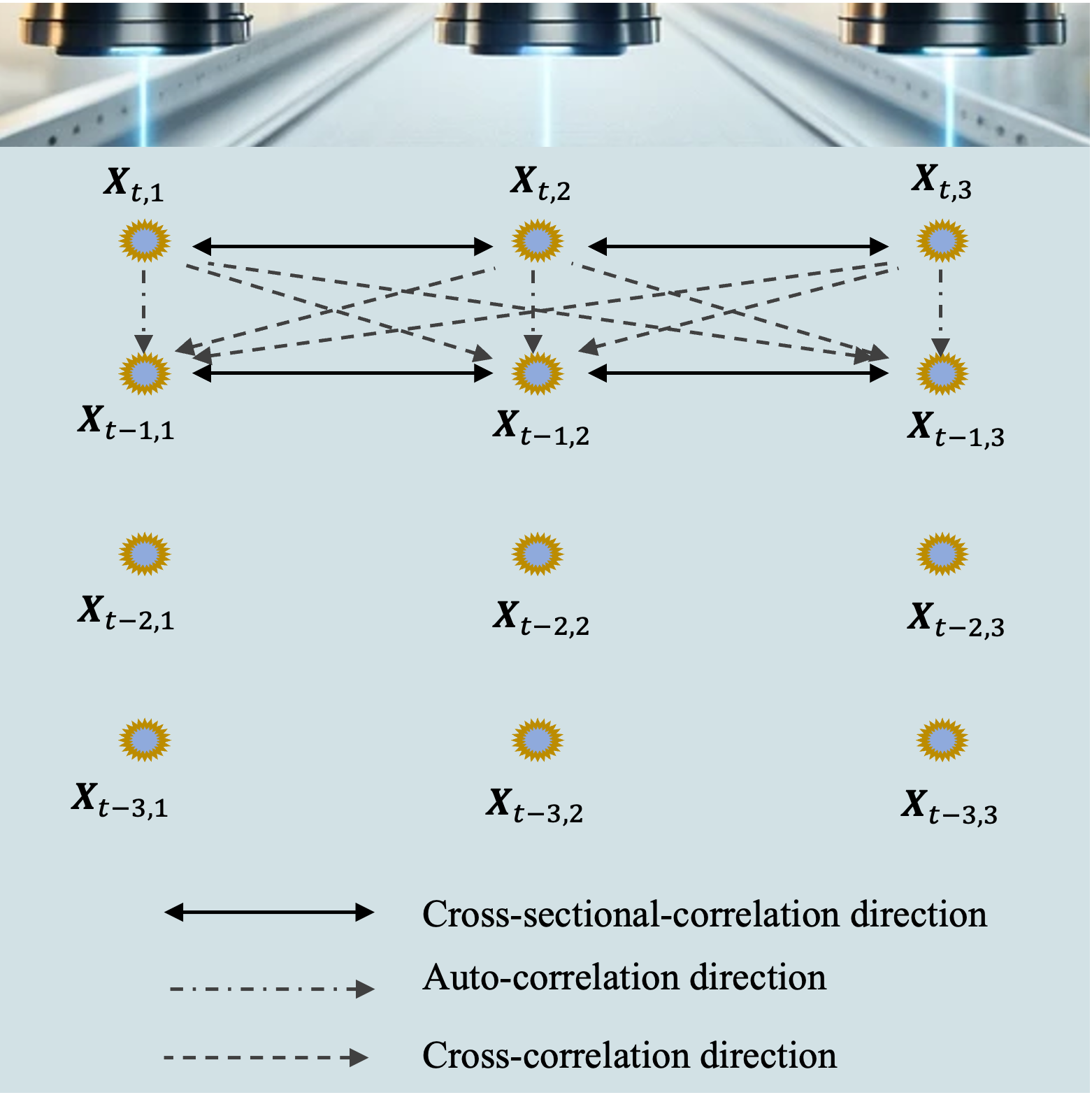}}}
 \caption{Schematic (generated by AI) of a steel sheet rolling process and a measured steel sheet.}
\label{fig:steelsheet}
\end{figure}

Nowadays, thanks to the rapid evolution of measuring equipment, the thickness of steel sheets can be measured by state-of-the-art non-contact measuring gauges like optical or ultrasonic-based thickness gauges (see Figure \ref{fig:steelsheet} (a)). A critical feature of such gauges is their ability to collect measurements in sheets frequently in time. Accordingly, moderate to high degrees of autocorrelation and cross-correlation are expected to appear among consecutive measurements (Sparks (2015)).

Consider a continuous steel sheet rolling process where measurements of thickness are collected at $v=3$ different cross-section locations. Each thickness reading is modelled as a normal random variable, and these measurements are cross-sectionally-correlated with one another. Therefore, an observation $\textbf{X}_t$ is a row of $v$ measurements (thickness in millimetres). For a sample of size $n$, the $n$ consecutive rows are grouped, and here, we assume that $n=4$. Figure \ref{fig:steelsheet}(b) illustrates the locations of the $v \times n=12$ thickness readings within a sample on the rolled steel sheet. Sparks (2015) stated that the adjacent measurements could be highly correlated and cross-correlated along rows and columns with degrees up to $\rho=0.95$.  Let us assume that a VAR(1) model is the time series model to fit the thickness measurements. For illustration, let us consider the following parameters as the IC model of the process. 
\begin{align}\label{equ:var1PhaseII}
\pmb{\mu}_{\textbf{X}}=(5,5,5)^{\intercal}, \quad \pmb{\Phi}=\begin{pmatrix}
0.70 & 0.15 & 0.15\\
0.15 & 0.70 & 0.15\\
0.15 & 0.15 & 0.70\\
\end{pmatrix}, \quad \pmb{\Sigma}_{\pmb{\epsilon}}=\begin{pmatrix}
1 & 0.70 & 0.70 \\
0.70 & 1 & 0.70\\
0.70 & 0.70  & 1
\end{pmatrix}.
\end{align}

The eigenvalues of $\pmb{\Phi}$ are $0.91,0.55,$ and $0.18$ showing that the trivariate time series would be stationary by these settings. Note that though the matrix $\pmb{\Phi}$ represents the linear dependence among observations, it is not an interpretable measure. To better measure this kind of relationship, we work with the cross-correlation. The above specific for the VAR(1) model and the fact that $\pmb{\rho}(k)=\textbf{D}^{-1}\pmb{\Gamma}(k)\textbf{D}^{-1}$ where $\pmb{\rho}(k)$ is the cross-correlation matrix at lag $k=0,1,\ldots$ and $\textbf{D}=diag(\sigma_{X_1},\ldots,\sigma_{X_v})$ give us the following cross-correlation matrices at lags $k=0,1,2$ 
\begin{align*}\label{rhomat}
\pmb{\rho}(0)&=\begin{pmatrix}
1.00 & 0.93 & 0.85\\
0.93 & 1.00 & 0.85\\
0.85 & 0.85 & 1.00\\
\end{pmatrix}, \qquad \pmb{\rho}(1)=\begin{pmatrix}
0.91 & 0.88 & 0.81 \\
0.88 & 0.91 & 0.81\\
0.71 & 0.71  & 0.69
\end{pmatrix}, \qquad \pmb{\rho}(2)=\begin{pmatrix}
0.52 & 0.49 & 0.45 \\
0.49 & 0.52 & 0.45\\
0.13 & 0.13  & 0.13
\end{pmatrix},
\end{align*}
that demonstrate high degrees of correlation and cross-correlation among data up to lag two. The cross-correlation matrices show that the linear dependency between observations tends to disappear if their observation times are far away. Given the parameters in \eqref{equ:var1PhaseII}, the  variance-covariance matrices of the quality variable $\textbf{X}_t$ and the sample mean $\overline{\textbf{X}}_t$ are obtained from \eqref{equ:VecSigmaW} and \eqref{equ:VCwbar} as:
\begin{align}
\pmb{\Sigma}_{\textbf{X}}=\begin{pmatrix}
6.39& 5.93 & 3.16\\
5.93& 6.36 & 3.16\\
3.16& 3.16 & 2.16
\end{pmatrix}, \quad \pmb{\Sigma}_{\overline{\textbf{X}}}=\begin{pmatrix}
5.70& 5.46 & 2.79\\
5.46& 5.70 & 2.79\\
2.79& 2.79 & 1.56
\end{pmatrix}.
\end{align}

\begin{figure}
\centering
 \subfigure[Time series plots of the IC and OC thickness measurements.]{%
 \resizebox*{7.5cm}{!}{\includegraphics{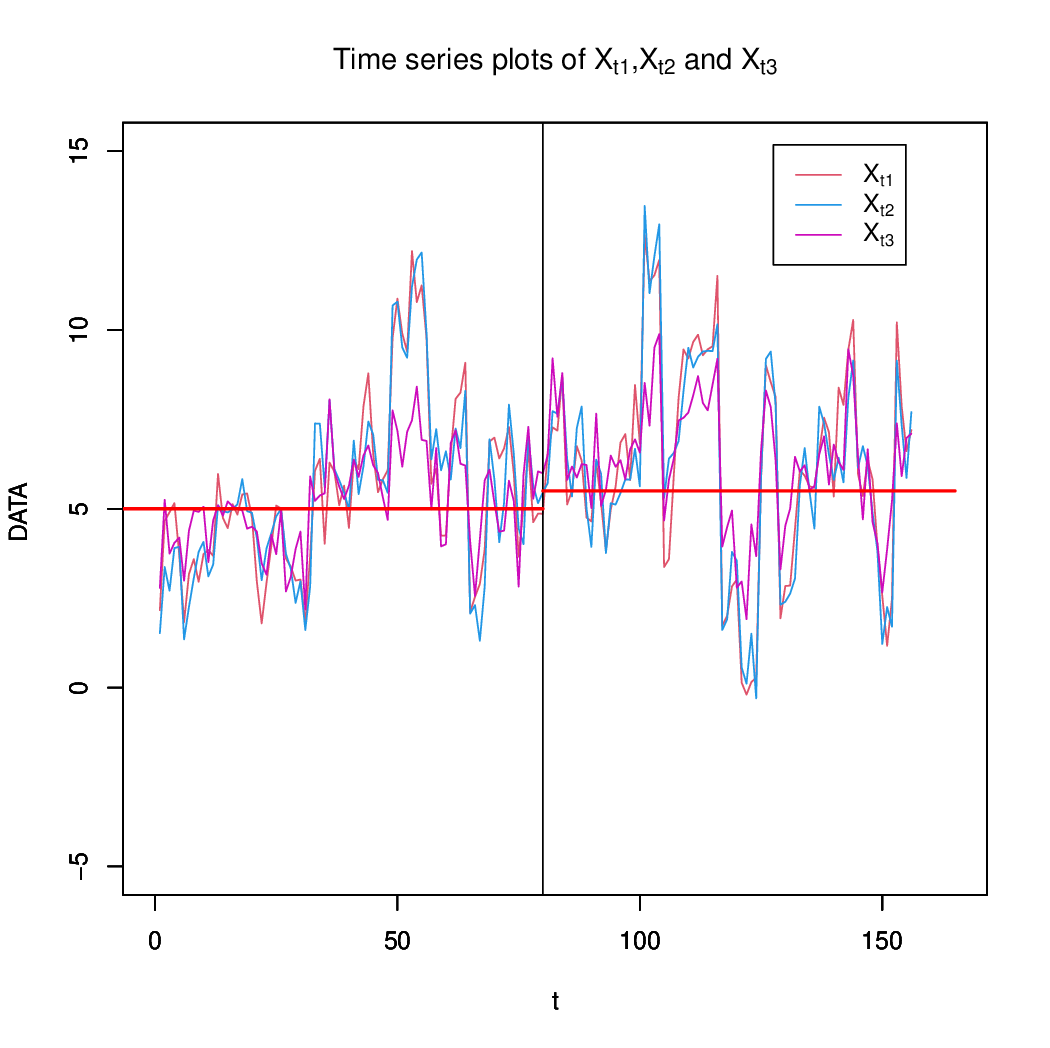}}}
 \subfigure[Hotelling's $T^2$ control chart based on   IC and OC thickness measurements.]{%
 \resizebox*{7.5cm}{!}{\includegraphics{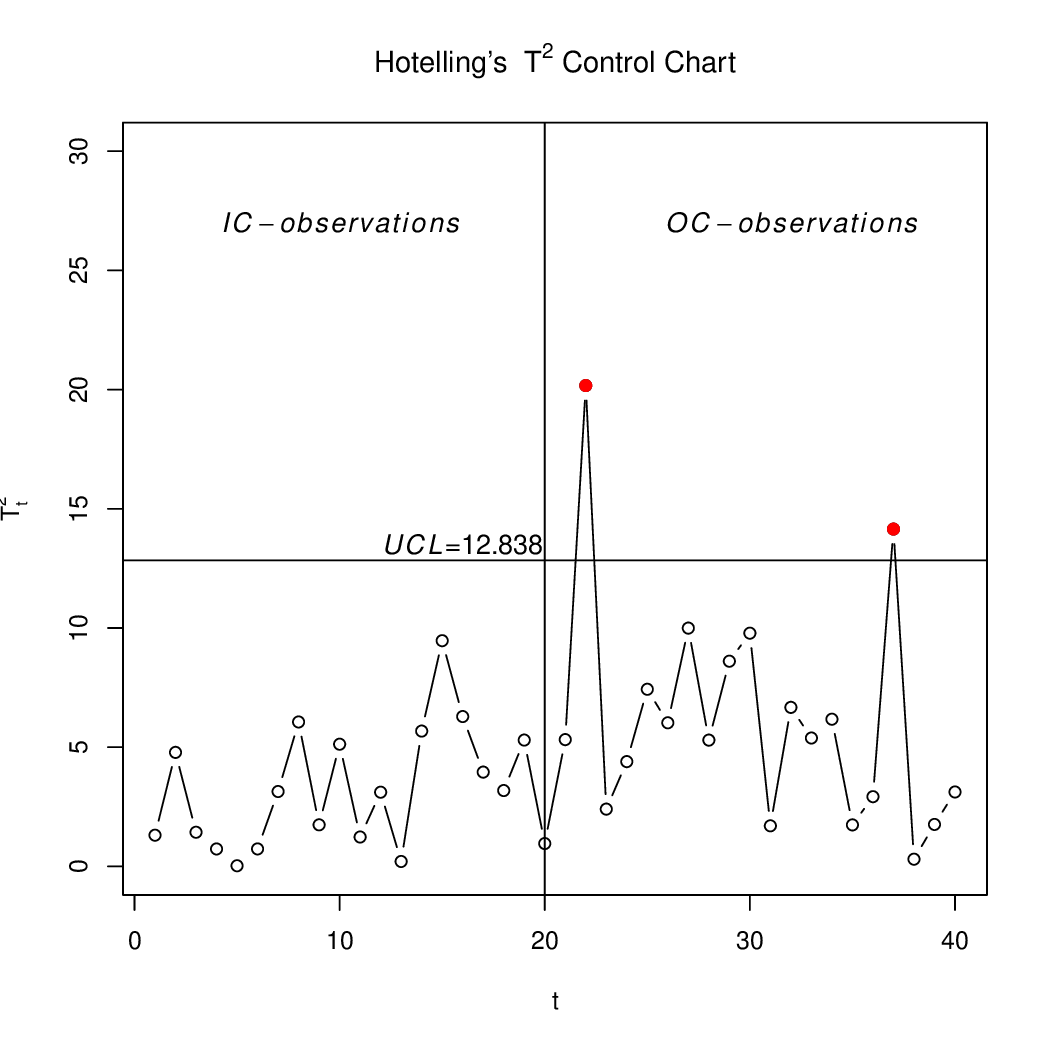}}}
 \caption{Time series plots and the Hotelling's $T^2$ control chart for the rolling steel sheets example.}
\label{fig:stellsheetT2chart}
\end{figure}

\subsubsection*{Phase II illustration}

In order to investigate the control chart's ability to detect the possible shifts in the mean vector of thickness, 20 samples of size $n=4$ were generated from a VAR(1) model with \emph{known} parameters, given in \eqref{equ:var1PhaseII}. If we set $ARL_0=200$, the $UCL$ of the $T^2$ control chart is $\chi^{2}_{3}(0.005)=12.838$. To simulate an OC situation, 20 samples were also generated from a VAR(1) model with given parameters in \eqref{equ:var1PhaseII} except for the shifted mean $\pmb{\mu}^1_{\textbf{X}}=\pmb{\mu}^0_{\textbf{X}}+0.50$.  Figure \ref{fig:stellsheetT2chart} (a) shows the plots of the three individual time series.  Figure \ref{fig:stellsheetT2chart}(b) illustrates that Hotelling $T^2$   control chart detected the mean shift at the $2nd$  and the $17^{th}$ OOC  samples after the process shift.

\subsection{Chemical process: VAR(3) observations}

There are various industries like food (Cullen et al. (2000)), polymer extrusion (Chen et al. (2003), and  Liu et al. (2010)), and chemical (Walz et al. (2014)) processes where the viscosity of a product is an important indicator of its quality. On the other hand, it is known that viscosity is highly affected by temperature. In liquids, viscosity usually decreases with increasing temperature, whereas, in most gases, it increases with temperature. So, taking into account the correlation between viscosity and temperature is of great importance in the design of any control chart for monitoring the quality of such processes. This leads us to use Hotelling's $T^2$ control chart to monitor the viscosity and temperature of a chemical process.

\begin{table}
 \begin{center}
 \caption{The viscosity $X_{t1}$ and temperature $X_{t2}$   Phase I observations.}
 \begin{tabular}{ccccccccccccccc}
 \hline
$t$ & $X_{1}$ & $X_{2}$ & & $t$  & $X_{1}$ & $X_{2}$ & & $t$  & $X_{1}$ & $X_{2}$ & & $t$  & $X_{1}$ & $X_{2}$\\
 \hline
 1 & 0.17 & 0.30 & & 26 & 0.08 & 0.53 & & 51 & 0.00 & 0.34 & & 76 &-0.04 &-0.12 \\
 2& 0.13 & 0.18 & & 27 & 0.17 & 0.54 & & 52 &0.02 &0.13 & & 77 & 0.11 & -0.26 \\
 3& 0.19 & 0.09 & & 28 &0.20 & 0.42 & & 53 & -0.08 & 0.21 & & 78 & 0.19 &0.20 \\
 4& 0.09 & 0.06 & & 29 &0.20 & 0.37 & & 54&-0.08 & 0.06 & & 79 & -0.07 & 0.18\\ 
 5& 0.03 & 0.30 & & 30 & 0.27 & 0.34 & & 55& -0.26 & 0.04 && 80 & -0.10 & 0.32 \\
 6& 0.11 & 0.44 & & 31 & 0.23 & 0.27 & & 56 & -0.06 & -0.06 & & 81 & -0.13 & 0.50\\
 7& 0.15 & 0.46 & & 32 & 0.23 & 0.34 & & 57& -0.06 & -0.16 & & 82 & 0.10 & 0.40 \\
 8& -0.02 & 0.44 & & 33 & 0.20 & 0.35 & & 58& -0.09 &-0.47 & & 83 &-0.10 & 0.41 \\
 9& 0.07 & 0.34 & & 34 & 0.08 & 0.43 & & 59&-0.14 & -0.50 & & 84 & -0.05 & 0.47 \\
 10& 0.00 & 0.23 & & 35 & -0.16 & 0.63 & & 60 & -0.10 & -0.60 & & 85 &-0.12 & 0.37 \\ 
 11& -0.08 & 0.07 & & 36 & -0.08 & 0.61 & & 61 & -0.25 &-0.49 & & 86 & 0.00 & 0.04 \\
 12& -0.15 & 0.21 & & 37 & 0.14 & 0.52 & & 62& -0.23 & -0.27 & & 87 & 0.03 & -0.10 \\
 13& -0.15 & 0.03 & & 38 & 0.17 & 0.06 & & 63& -0.11 & -0.18 & & 88 & -0.06 & -0.34 \\
 14& 0.04 & -0.20 & & 39 &0.27 & -0.11 & & 64 &-0.01 & -0.37 & & 89 & 0.03 &-0.41 \\
 15& 0.08& -0.39 & & 40 & 0.19 & -0.01 & & 65 &-0.17 & -0.34 & & 90 & 0.04 &-0.33 \\
 16& 0.10 & -0.70 & & 41 & 0.10 & 0.02 & & 66 & -0.23 & -0.34 & & 91 & 0.09 &-0.25 \\
 17& 0.07 &-0.22 & & 42 & 0.13 &0.34 & & 67 & -0.28 & -0.18 & & 92 & -0.25 &-0.18 \\
 18&-0.01 & -0.08 & & 43 & -0.05 & 0.21 & & 68 &-0.26 & -0.26 & & 93 & -0.25 & -0.06 \\
 19& 0.06 & 0.16 & & 44 & 0.13 & 0.18 & & 69 &-0.19 & -0.51 & & 94 & -0.40 & 0.15 \\
 20& 0.07 & 0.13 & & 45 & -0.02 & 0.19 & & 70 & -0.26 & -0.65 & & 95 & -0.30 & -0.32 \\
 21& 0.17 & 0.07 & & 46 & 0.04 & 0.05 & & 71& -0.20 & -0.71 & & 96 & -0.18 & -0.32 \\
 22& -0.01 & 0.23 & & 47& 0.00 & 0.15 & & 72 & -0.08 & -0.82 & & 97 & -0.09 & -0.81 \\
 23& 0.09 & 0.33 & & 48 & 0.08 & 0.10 & & 73& 0.03 & -0.70 & & 98 & -0.05 & -0.87 \\
 24& 0.22 & 0.72 & & 49 & 0.08 & 0.28 & & 74&-0.08 & -0.63 & & 99 & 0.09 &-0.84 \\
 25& 0.09 & 0.45 & & 50 & 0.07 & 0.20 & & 75 & -0.01 & -0.29 & & 100 & 0.18 &-0.73 \\
 \hline
 \end{tabular}
  \label{tab:phaseIchemicaldata}
 \end{center}
\end{table}

 \subsubsection*{Phase I implementation}

\begin{figure}
\centering
 \subfigure[ Time series plots.]{%
 \resizebox*{4cm}{!}{\includegraphics{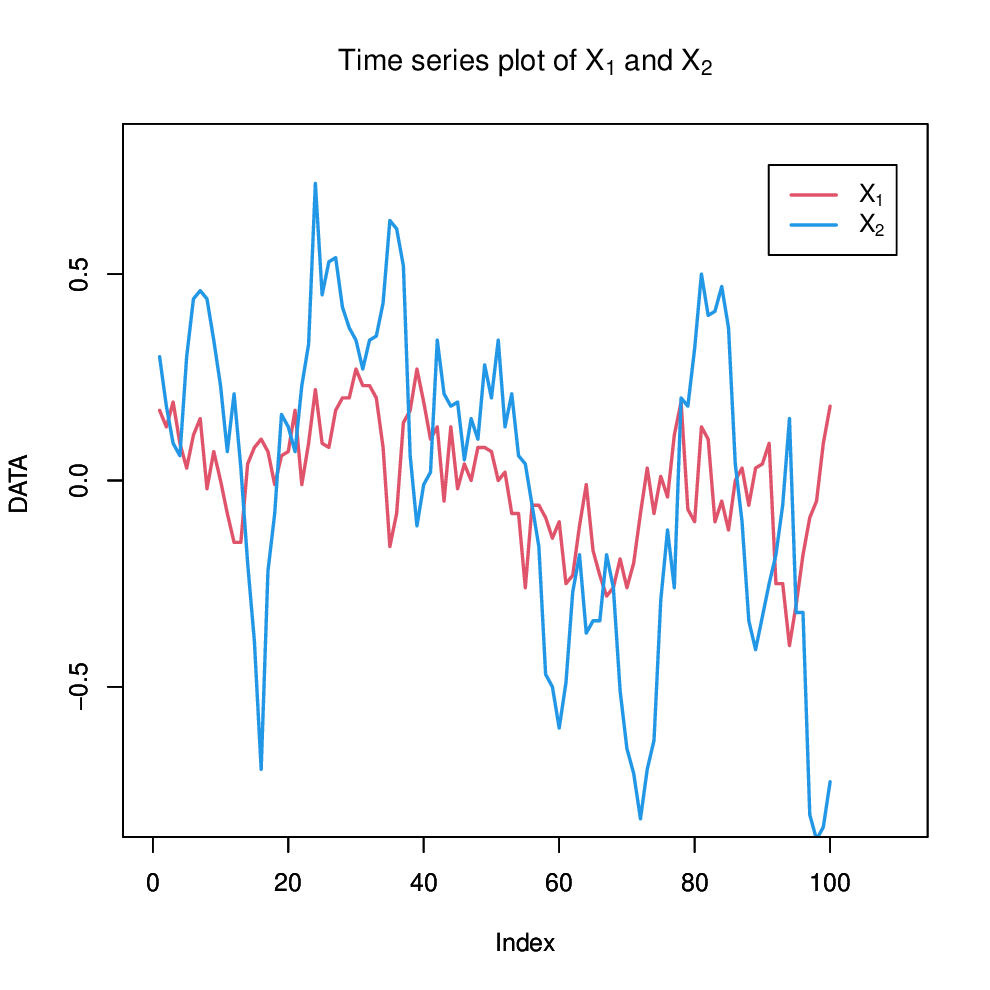}}}
 %\subfigure[QQ plot  for $X_1$.]{%
 %\resizebox*{4cm}{!}{\includegraphics{qqpoltx1}}}
 %\subfigure[QQ plot for $X_2$.]{%
 %\resizebox*{4cm}{!}{\includegraphics{qqpoltx2}}}
 \subfigure[ACF plot for $X_{t1}$.]{%
 \resizebox*{4cm}{!}{\includegraphics{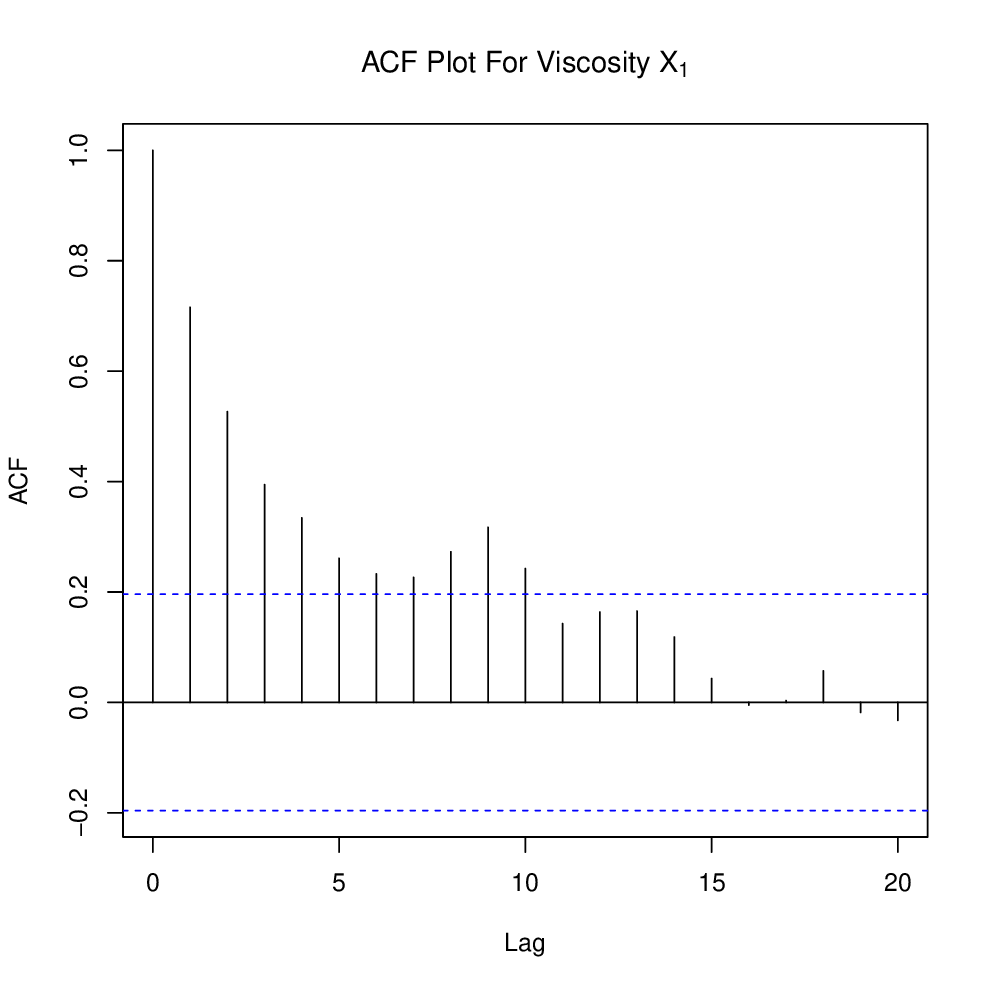}}}
 \subfigure[PACF plot for $X_{t1}$.]{%
 \resizebox*{4cm}{!}{\includegraphics{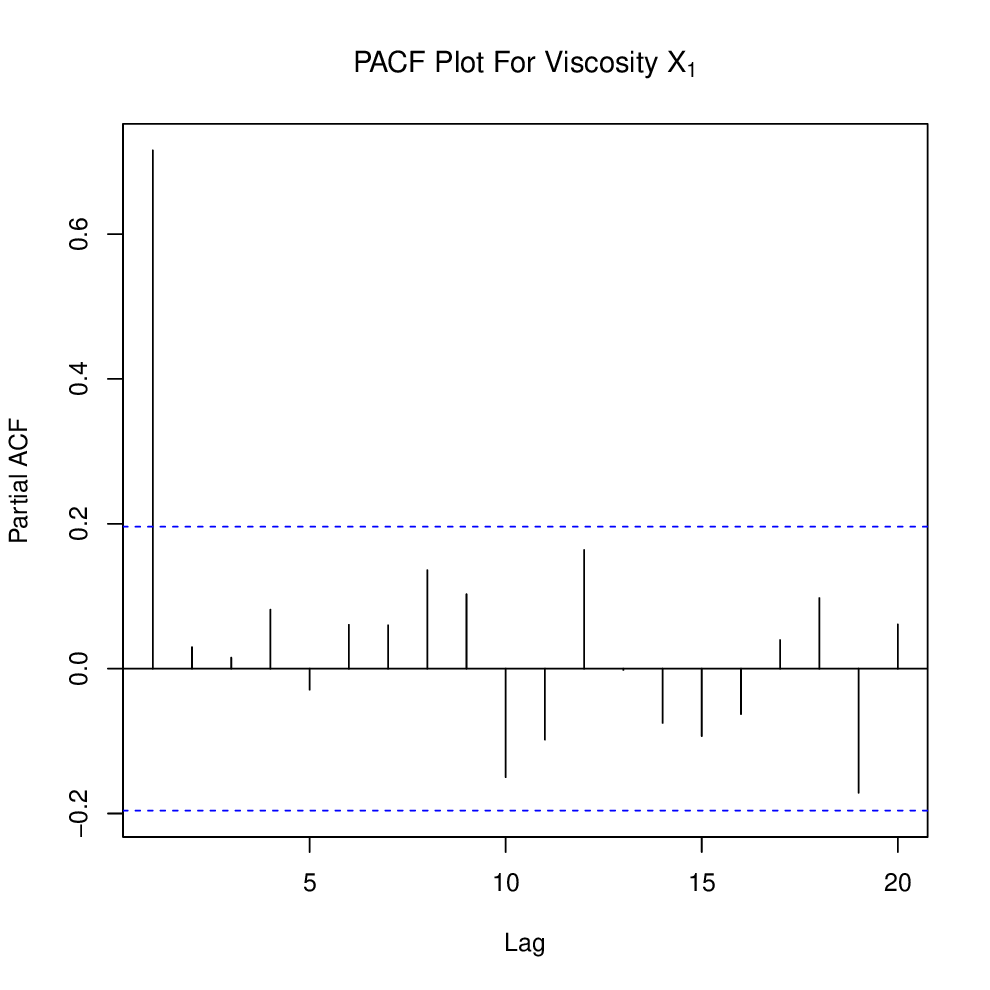}}} 
 \subfigure[ACF plot for $X_{t2}$.]{%
 \resizebox*{4cm}{!}{\includegraphics{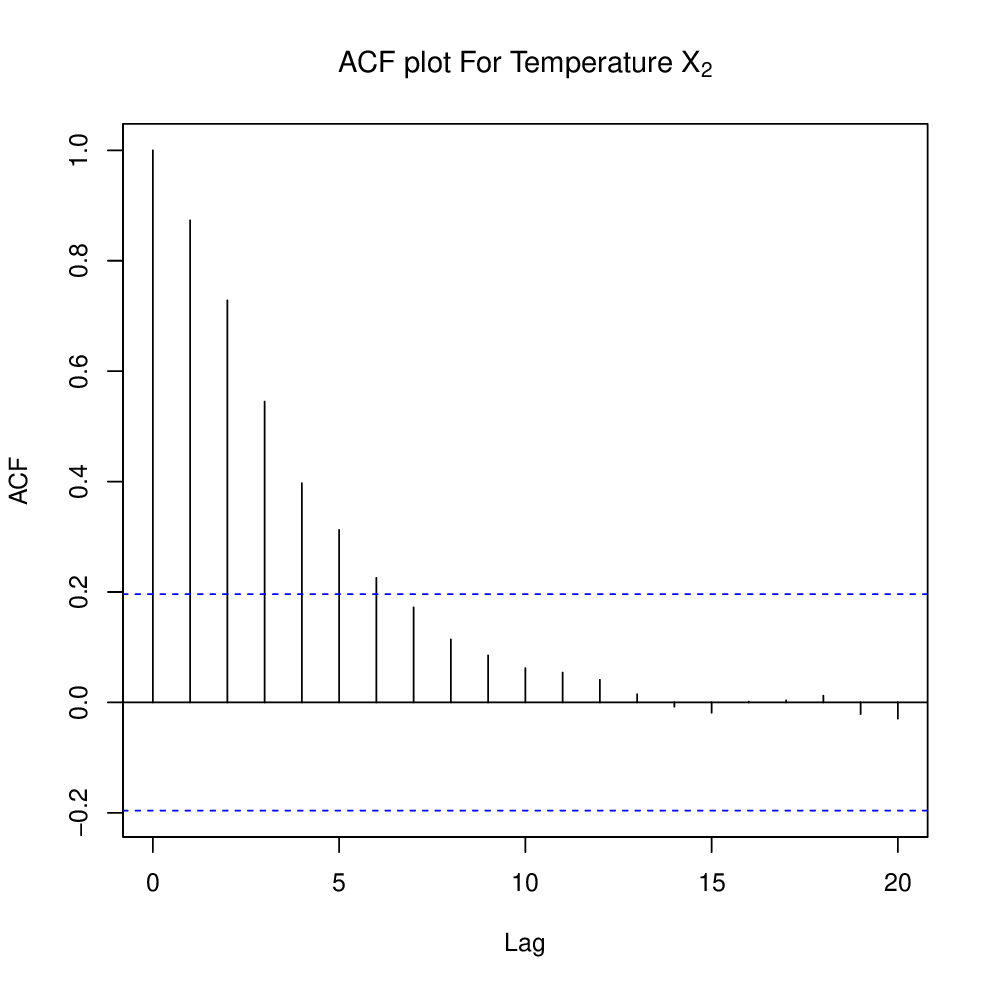}}}
\subfigure[ PACF plot for $X_{t2}$.]{%
 \resizebox*{4cm}{!}{\includegraphics{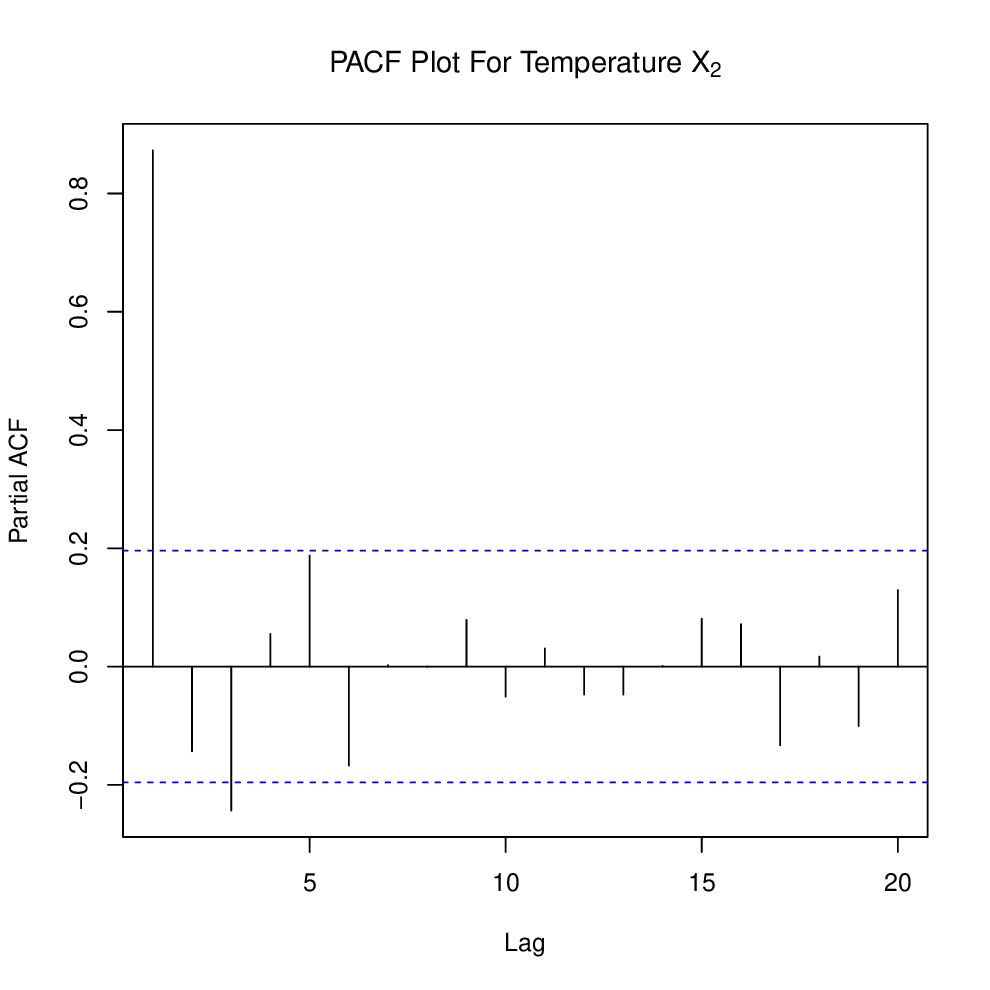}}}
 \subfigure[CCF plot for $X_{t1}$ and $X_{t2}$.]{%
 \resizebox*{4cm}{!}{\includegraphics{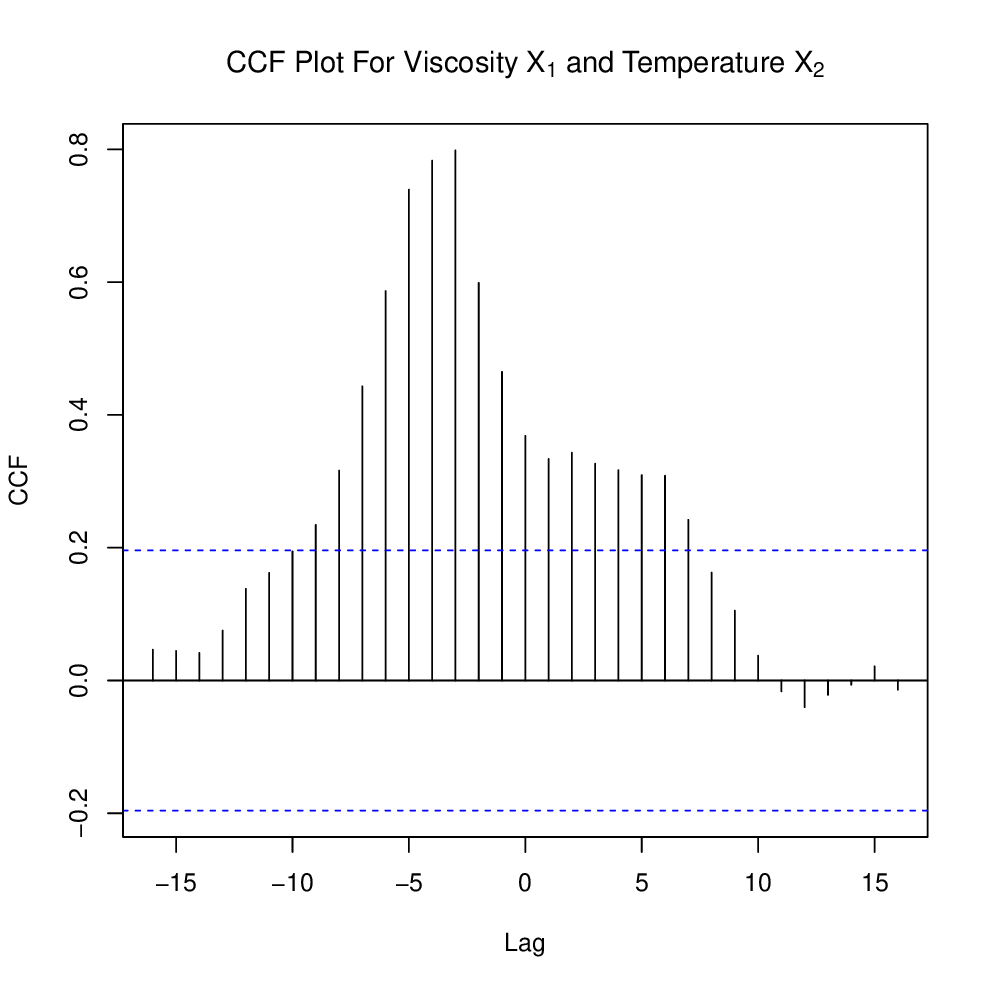}}}
%\subfigure[QQ plot  for $X_1$ Residuals.]{%
 %\resizebox*{4cm}{!}{\includegraphics{qqpoltx1res}}} 
%\subfigure[QQ plot  for $X_2$ Residuals.]{%
 %\resizebox*{4cm}{!}{\includegraphics{qqpoltx1res}}}\hfill
 \subfigure[ACF plot for   $X_{t1}$ Residuals.]{%
 \resizebox*{4cm}{!}{\includegraphics{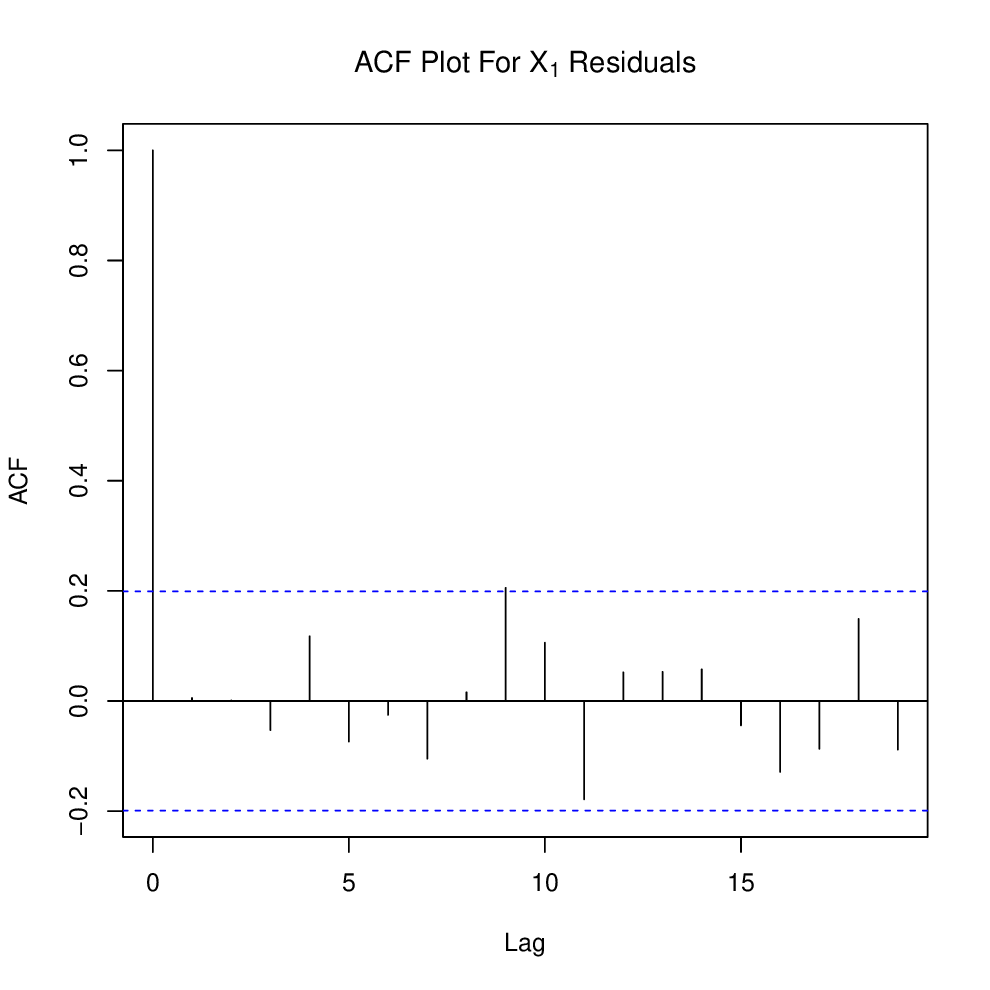}}}
\subfigure[ACF plot for   $X_{t2}$ Residuals.]{%
 \resizebox*{4cm}{!}{\includegraphics{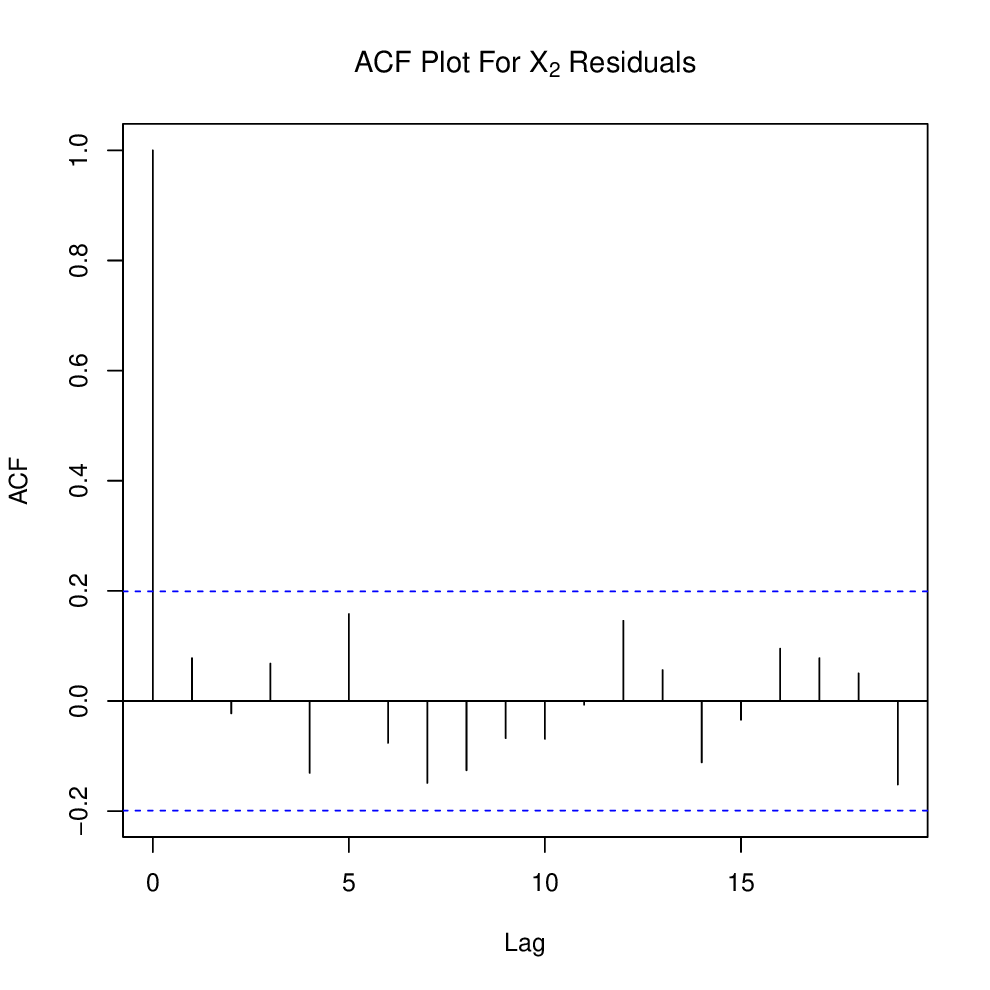}}} 
 \caption{Time series analysis of the chemical process dataset.}
\label{fig:timeseriesanalysis}
\end{figure}

Montgomery et al. (2015) provided a dataset of $m=100$ bivariate ($v=2$) individual ($n=1$) observations regarding the viscosity ($X_1$) and temperature ($X_2$) in a chemical process. The time-ordered data are presented in Table \ref{tab:phaseIchemicaldata}. The data set has been analyzed by Montgomery et al. (2015) in a time series analysis context, but we assessed the data in an SPM context to monitor the viscosity and temperature of the process output simultaneously. It is worth mentioning that the data are centred by subtracting the respective averages. Since the data are collected at high frequency, we expect the variables to exhibit some autocorrelation and cross-correlation. In order to implement the $T^2$ control chart, let us consider these observations as the Phase I dataset to estimate the model's parameters.  We used the \texttt{vars} package  (Pfaff (2008)) in \texttt{R} environment to get the numerical results.  Figure \ref{fig:timeseriesanalysis} (a) displays the univariate time series $X_{t1}$ and $X_{t2}$. The figure also shows that the temperature time series exhibits greater variability compared to the viscosity. Let us begin by checking the $p$-value of the Durbin-Watson approach to test for autocorrelation of data. The $p$-values for both individual time series $X_{t1}$ and $X_{t2}$ are less than $0.0001$ showing strong evidence in data against the null assumption $H_0$: no serial correlation. We continue by checking the stationarity of time series for $t=1,\ldots,100$. To do this, we applied the Dickey-Fuller (DF) method to test $H_0$: the time series is non-stationary. The $p$-value of the DF test for both time series is $< 0.01$, which means that it is reasonable to assume that both time series $X_{t1}$ and $X_{t2}$ are stationary. Testing the normality of observations with the \texttt{R} function \texttt{normality.test}, the  $p$-value of the multivariate Jarque-Bera test is equal to $0.2685$. So, it can be concluded that the Multivariate Normal distribution is a reasonable fit for the time series. To determine the autocorrelation structure of the time series, Figures \ref{fig:timeseriesanalysis}(b)-(f) display plots of the autocorrelation function (ACF), the partial autocorrelation function (PACF), and the cross-correlation function (CCF) for both time series. These figures demonstrate that: (i) in both cases an AR model seems to be suitable (see \ref{fig:timeseriesanalysis} Figures(b) and (cd)); (ii)  AR models of orders one and three seem to be appropriate for the time series $X_{t1}$ and $X_{t2}$, respectively (see Figures \ref{fig:timeseriesanalysis}(c) and (e)); (iii) there is a significant positive cross-correlation between the observations of $X_{t1}$ and $X_{t2}$ up to lag  $k=\pm 10$  (see Figure \ref{fig:timeseriesanalysis}(f)).  All these results demonstrate that a bivariate VAR model is an appropriate choice to describe the output of this process. In order to select the order $p$ of the VAR model, the \texttt{R}  function \texttt{VARselect} is used for $p$ up to 3. Using the Akaike information criterion (\textit{AIC}) metric that is obtained for the three candidate models as $-8.0276, -8.1059,$ and $-8.8202$,   the VAR(3) model can be selected as the right choice. Accordingly,  we fitted a VAR(3) model to the data using the \texttt{VAR} function. The estimated VAR(3) is obtained with mean $\pmb{\mu}_{\textbf{X}}=(0,0)^{\intercal}$:
\begin{align}\label{equ:modelparameters}
	\begin{pmatrix}
		X_{t,1}\\
		X_{t,2}
	\end{pmatrix}&=
\underbrace{\begin{pmatrix}
		-0.002\\
		-0.001
	\end{pmatrix}}_{\textbf{C}}
+\underbrace{\begin{pmatrix}
		0.690 & -0.043\\
		0.049 &  0.634
	\end{pmatrix}}_{\pmb{\Phi}_{1}}
 \begin{pmatrix}
		X_{t-1,1}\\
		X_{t-1,2}
	\end{pmatrix}+\underbrace{\begin{pmatrix}
		0.010 & 0.091\\
		-0.016 &  0.270
	\end{pmatrix}}_{\pmb{\Phi}_{2}}
 \begin{pmatrix}
		X_{t-2,1}\\
		X_{t-2,2}
	\end{pmatrix}\nonumber\\
&+\underbrace{\begin{pmatrix}
		-0.006 & -0.017\\
		1.125 &  -0.317
	\end{pmatrix}}_{\pmb{\Phi}_{3}}
 \begin{pmatrix}
		X_{t-3,1}\\
		X_{t-3,2}
	\end{pmatrix}+\begin{pmatrix}
		\varepsilon_{X_{t,1}}\\
		\varepsilon_{X_{t,2}}
	\end{pmatrix},
\end{align} 
 where
 \begin{align}\label{equ:modelparameterss}
\begin{pmatrix}
		\varepsilon_{X_{t,1}}\\
		\varepsilon_{X_{t,2}}
	\end{pmatrix} \sim \textbf{MN}_2 \left(\textbf{0},\boldsymbol{\Sigma}_{\boldsymbol{\varepsilon}}=\begin{pmatrix}
		0.011&-0.001\\
		-0.001& 0.012
	\end{pmatrix}\right).
 \end{align}

 The estimated matrices $\pmb{\Phi}_{1},\pmb{\Phi}_{2},$ and $\pmb{\Phi}_{3}$ in \eqref{equ:modelparameters} and plots in Figure \ref{fig:timeseriesanalysis}(b)-(f) provide consistent information about the presence of autocorrelation and cross-correlation among observations up to lag $k=3$. Now, let us now to analyze the residuals.  QQ plots (not presented here) demonstrate the normality of the residuals, as well as the ACF plots (Figure \ref{fig:timeseriesanalysis} (g) and (h)) and the CCF plot (not presented here), show the absence of correlation and cross-correlation among residuals of the fitted VAR(3) model residuals.  Therefore, we conclude that the VAR(3) model provides a reasonable fit for the process, and the estimated parameters can be used for the online monitoring of the process mean vector in Phase II. However, the stability of the process mean in Phase I should be checked through a Hotelling $T^2$ control chart.

To carry out the Phase I   implementation of the  Hotelling $T^2$ control chart, let us consider that samples of size $n=5$ are taken from the process at each inspection. Thus, we grouped the original dataset into $20$ samples of size $n=5$. The sample observations, along with the sample mean and the $T^2$ values, are presented in Table \ref{tab:phaseIchemicalprocess}.  In order to calculate the monitored statistics $T_i^2$, we need to find the variance-covariance matrix of the sample mean $\pmb{\Sigma}_{\overline{\textbf{X}}}$ with the procedure shown in Section 2. Do this end, let us define the matrices $\pmb{\Psi}$ and $\pmb{\Sigma}_b$ according to \eqref{equ:newcoeff} as:
\begin{align*}
\pmb{\Psi}=\begin{pmatrix}
0.690 &   -0.043 &    0.010 &  0.091 &   -0.006 & -0.017 \\
0.049 &   0.633 & -0.016 &   0.270 &   1.125 & -0.317\\
1 &   0 & 0 &   0 &   0 & 0\\
0 &   1 & 0 &   0 &   0 & 0\\
0 &   0 & 1 &   0 &   0 & 0\\
0 &   0 & 0 &   1 &   0 & 0\\
\end{pmatrix},  
\pmb{\Sigma}_b=\begin{pmatrix}
0.011 & -0.001 & 0 &0 &0 &0 \\
-0.001 &0.012  &  0&0& 0&0 \\
0 &0 &  0&0& 0&0 \\
0 &0 &  0&0& 0&0 \\
0 &0 &  0&0& 0&0 \\
0 &0 &  0&0& 0&0 \\
\end{pmatrix}.
\end{align*}

We also assessed the fitted VAR(3) stationarity by checking the eigenvalues of $\pmb{\Psi}$, all of which were found to be less than 1 in absolute value. Using \eqref{equ:VecSigmaZ}, the variance-covariance matrix of the expanded matrix $\textbf{Z}_t$ is obtained as:
 \begin{align*}
\pmb{\Sigma}_{\textbf{Z}}=\begin{pmatrix}
\underbrace{\begin{pmatrix}
		0.023 &  0.020\\
		0.020 &  0.165
	\end{pmatrix}}_{\pmb{\Sigma}_{\textbf{X}}}
 & \underbrace{\begin{pmatrix}
		0.016 &  0.018\\
		0.026 &  0.146
	\end{pmatrix}}_{\pmb{\Gamma}(1)} & \underbrace{\begin{pmatrix}
		0.012 &  0.019\\
		0.035 &  0.120
	\end{pmatrix}}_{\pmb{\Gamma}(2)}\\
\underbrace{\begin{pmatrix}
		0.016 &  0.026\\
		0.018 &  0.146
	\end{pmatrix}}_{\pmb{\Gamma}(-1)}
 & \underbrace{\begin{pmatrix}
		0.023 &  0.020\\
		0.020 &  0.165
	\end{pmatrix}}_{\pmb{\Sigma}_{\textbf{X}}} &
 \underbrace{\begin{pmatrix}
		0.016 &  0.018\\
		0.026 &  0.146
	\end{pmatrix}}_{\pmb{\Gamma}(1)}\\
\underbrace{\begin{pmatrix}
		0.012 &  0.035\\
		0.019 &  0.120
	\end{pmatrix}}_{\pmb{\Gamma}(-2)}
 & \underbrace{\begin{pmatrix}
		0.016 &  0.026\\
		0.018 &  0.146
	\end{pmatrix}}_{\pmb{\Gamma}(-1)} &
 \underbrace{\begin{pmatrix}
		0.023 &  0.020\\
		0.020 &  0.165
	\end{pmatrix}}_{\pmb{\Sigma}_{\textbf{X}}}\\
\end{pmatrix}.
\end{align*}

Now, based on equation \eqref{equ:VCwbar}, we have:
  \begin{align*}
\pmb{\Sigma}_{\overline{\textbf{Z}}}=\begin{pmatrix}
\underbrace{\begin{pmatrix}
		0.015  & 0.026  \\
		 0.026 &   0.127
	\end{pmatrix}}_{\pmb{\Sigma}_{\overline{\textbf{X}}}}
 & \underbrace{\begin{pmatrix}
		0.014  &  0.022 \\
		0.030  &   0.123
	\end{pmatrix}}_{\pmb{\Gamma}_{\overline{\textbf{X}}}(1)}
 & 
 \underbrace{\begin{pmatrix}
		 0.013 & 0.019  \\
		0.035  &   0.110
	\end{pmatrix}}_{\pmb{\Gamma}_{\overline{\textbf{X}}}(2)}\\
\underbrace{\begin{pmatrix}
		0.014  &0.030   \\
		 0.022 &   0.123
	\end{pmatrix}}_{\pmb{\Gamma}_{\overline{\textbf{X}}}(-1)}
 & 
 \underbrace{\begin{pmatrix}
		0.015  &   0.026 \\
		0.026  &   0.127
	\end{pmatrix}}_{\pmb{\Sigma}_{\overline{\textbf{X}}}}
 & 
 \underbrace{\begin{pmatrix}
		 0.014 & 0.022  \\
		  0.030 & 0.123  
	\end{pmatrix}}_{\pmb{\Gamma}_{\overline{\textbf{X}}}(1)}\\
\underbrace{\begin{pmatrix}
		0.013  &  0.035 \\
		0.019  &   0.110
	\end{pmatrix}}_{\pmb{\Gamma}_{\overline{\textbf{X}}}(-2)}
 & 
 \underbrace{\begin{pmatrix}
		0.014  &  0.030  \\
		0.022  &    0.123
	\end{pmatrix}}_{\pmb{\Gamma}_{\overline{\textbf{X}}}(-1)}
 & 
 \underbrace{\begin{pmatrix}
		0.015  & 0.026  \\
		0.026  &   0.127
	\end{pmatrix}}_{\pmb{\Sigma}_{\overline{\textbf{X}}}}
\end{pmatrix}
\end{align*}

\begin{table}
\caption{The Viscosity $X_1$ and Temperature $X_2$   of the chemical process for Phase I observations.}
\setlength{\tabcolsep}{15pt}
\renewcommand{\arraystretch}{0.6}
	\label{tab:phaseIchemicalprocess}
	\begin{center}
			\begin{tabular}{c|ccccc|cc}
				\hline
				 $t$ & $X_{t,1,1}$ &$X_{t,1,2}$ &$X_{t,1,3}$ &$X_{t,1,4}$ & $X_{t,1,5}$     & $\overline{X}_{t,1}$  &   $T_t^2$  \\
				&  $X_{t,2,1}$ & $X_{t,2,2}$ &$X_{t,2,3}$  &$X_{t,2,4}$  & $X_{t,2,5}$     & $\overline{X}_{t,2}$   &\\
				\hline
				 1  &   0.17   & 0.13  & 0.19  & 0.09 & 0.03  & 0.122 &   1.025 \\
				&  0.30& 0.18 & 0.09  & 0.06   &0.30  & 0.186 & \\[2mm] \hline
				 2    &  0.11 & 0.15 & -0.02  &  0.07 & 0.00 &  0.062&   1.168 \\
				& 0.44 & 0.46  & 0.44  &0.34 &0.23  &0.382  & \\[2mm]\hline
				 3  &   -0.08   &-0.15  & -0.15   &0.04  & 0.08  &-0.052  & 0.199   \\
				&0.07  &0.21  & 0.03  &-0.20    &-0.39  & -0.056& \\[2mm]\hline
				  4   & 0.10  & 0.07 & -0.01  &  0.06 & 0.07 & 0.058  &  0.949 \\
				& -0.70 & -0.22  & -0.08  &0.16 & 0.13 & -0.142 & \\[2mm]\hline
				5  &  0.17    & -0.01  & 0.09  &0.22  &  0.09 &  0.112 & 1.181   \\
				&  0.07& 0.23 &  0.33 & 0.72   &0.45  &  0.360 & \\[2mm]\hline
				 6    &0.08  &0.17  &0.20   & 0.20  & 0.27 &  0.184 & 2.478  \\
				& 0.53 & 0.54  & 0.42  &0.37 & 0.34 & 0.440& \\[2mm]\hline
				7  &   0.23   &0.23   & 0.20  & 0.08 & -0.16  & 0.116  & 1.407   \\
				&  0.27& 0.34 & 0.35  & 0.43   &0.63  & 0.404 & \\[2mm]\hline
				 8   & -0.08  & 0.14 & 0.17  &  0.27 & 0.19 & 0.138 &   1.308 \\
				& 0.61 & 0.52  & 0.06  &-0.11 &-0.01  & 0.214 & \\[2mm]\hline
				9  &  0.10    &0.13   & -0.05  & 0.13 & -0.02  & 0.058 &  0.320  \\
				&0.02  &0.34  &0.21   & 0.18   & 0.19 & 0.188 & \\[2mm]\hline
				 10   &  0.04 & 0.00 & 0.08  &  0.08 & 0.07 & 0.054  &  0.245   \\
				& 0.05 & 0.15  & 0.10  &0.28 & 0.20 & 0.156& \\[2mm]\hline
				11 &    0.00  &0.02   &  -0.08 & -0.08 &-0.26   & -0.080 & 1.499   \\
				& 0.34 & 0.13 & 0.21  & 0.06   &0.04  &0.156 & \\[2mm]\hline
				 12 &  -0.06 & -0.06 & -0.09  & -0.14  &-0.10  & -0.090 &  1.039  \\
				& -0.06 & -0.16  &-0.47   &-0.50 & -0.60 &-0.358 & \\[2mm]\hline
				13  &   -0.25   &-0.23   &-0.11   &  -0.01&  -0.17  & -0.154 &  1.662  \\
				&-0.49  & -0.27 &-0.18   &-0.37    &-0.34  & -0.330 & \\[2mm]\hline
				  14    & -0.23  & -0.28 & -0.26  & -0.19  &-0.26  & -0.244 &  4.080  \\
				& -0.34 & -0.18  & -0.26  & -0.51& -0.65 & -0.388 & \\[2mm]\hline
				15  &     -0.20 & -0.08  &  0.03 & -0.08 & -0.01  & -0.064 & 3.531   \\
				& -0.71 & -0.82 &-0.70   & -0.63   &-0.29  & -0.630 & \\[2mm]\hline
				 16   &  -0.04 &0.11  & 0.19  &-0.07   &-0.10  &0.018   &  0.035 \\
				& -0.12 & -0.26  & 0.20  &0.18 & 0.32 &  0.064 & \\[2mm]\hline
				17  &    -0.13  & -0.10  & 0.10  & -0.05 & -0.12  & -0.008 &  2.394  \\
				&  0.50& 0.40 & 0.41  & 0.47   &0.37  & 0.430 & \\[2mm]\hline
				18  & 0.00  &0.03  & -0.06  & 0.03  & 0.04 &  0.008  & 0.714 \\
				& 0.04 & -0.10  & -0.34  &-0.41 & -0.33 &-0.228 & \\[2mm]\hline
				19  &   0.09   &-0.25   &-0.25   & -0.40 & -0.30  & -0.222 &   4.161 \\
				& -0.25 & -0.18 &-0.06   &0.15    &-0.32  & -0.132 & \\[2mm]\hline
				20   & -0.18  &-0.09  & -0.05  & 0.09  & 0.18 & 0.033 &   9.234 \\
				&-0.32  &-0.81   &-0.87   &-0.84 & -0.73 &-0.812 & \\[2mm]
				\hline
		\end{tabular}
	\end{center}
\end{table}

\begin{figure}
 \centering
 \subfigure[ACF   plot  of $T_t^2$ values.]{%
 \resizebox*{7.5cm}{!}{\includegraphics{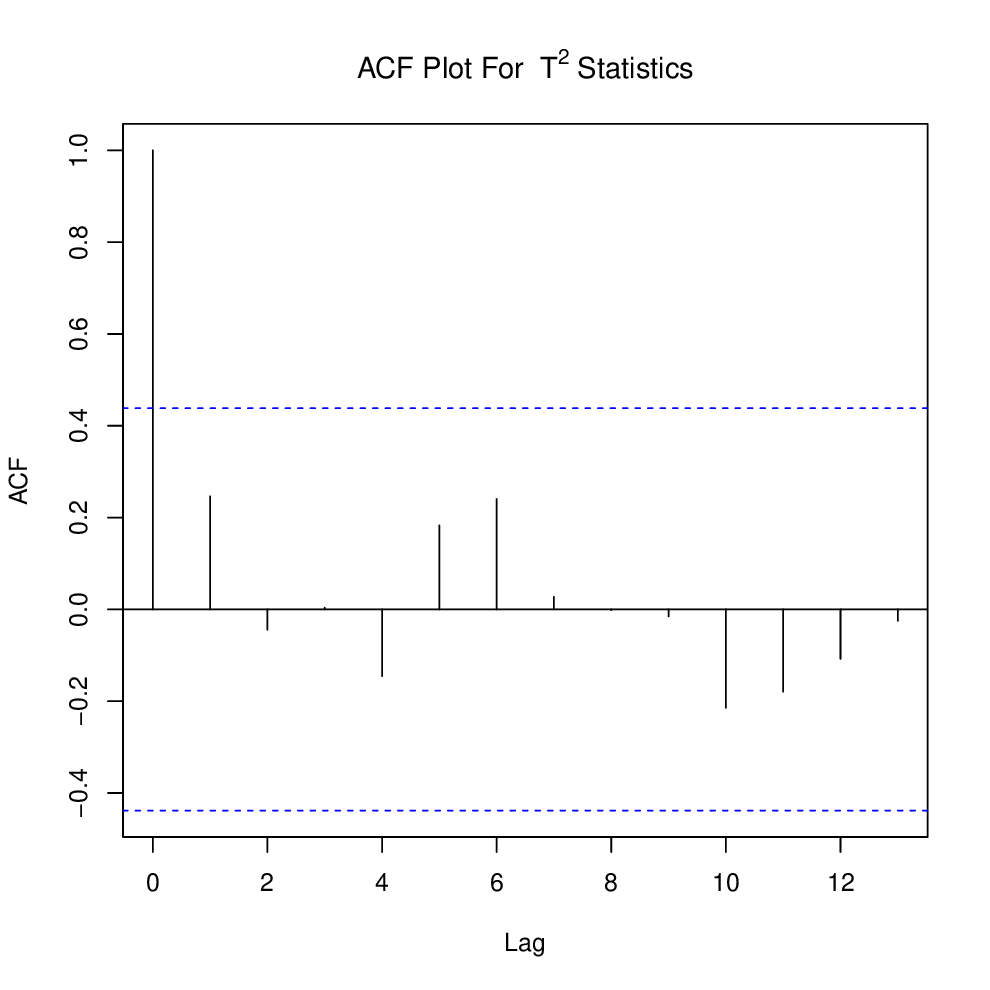}}}
 \subfigure[PACF   plot  of $T_t^2$ values.]{%
 \resizebox*{7.5cm}{!}{\includegraphics{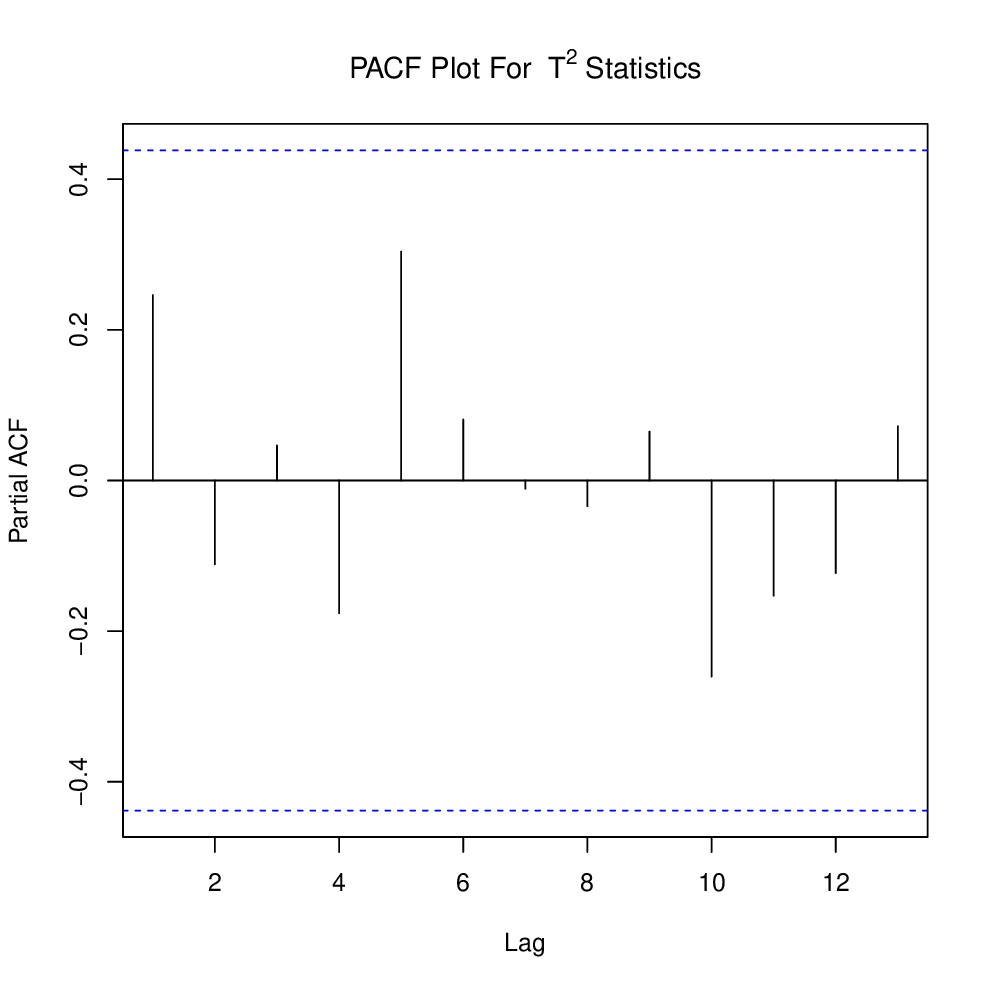}}}
 \caption{Sample ACF and PACF plots of $T_t^2$ statistics when $n=5$ in the chemical process example.}
\label{fig:ACFT2}
\end{figure}

\begin{figure}
\centering
 \subfigure[Phase I control chart.]{%
 \resizebox*{7.5cm}{!}{\includegraphics{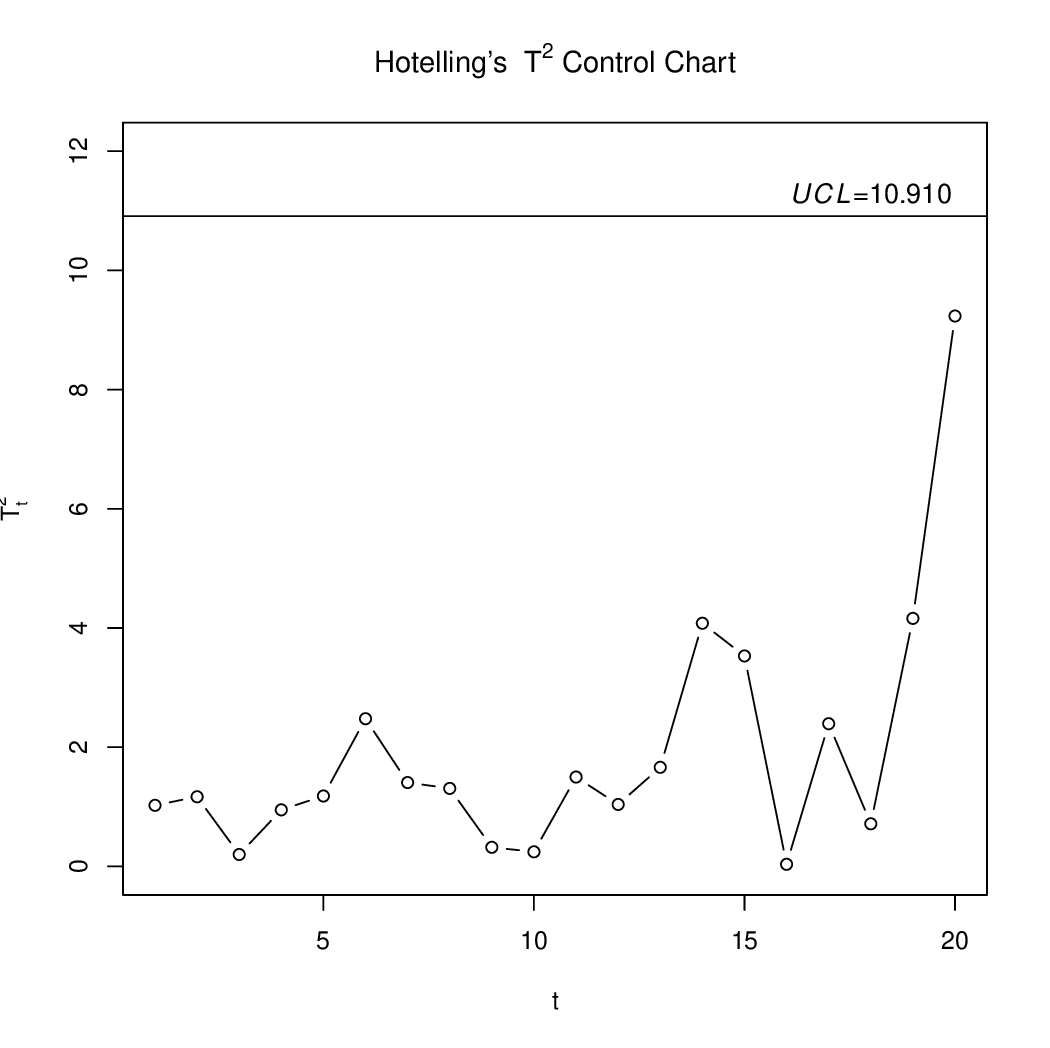}}}
 \subfigure[Phase II control chart.]{%
 \resizebox*{7.5cm}{!}{\includegraphics{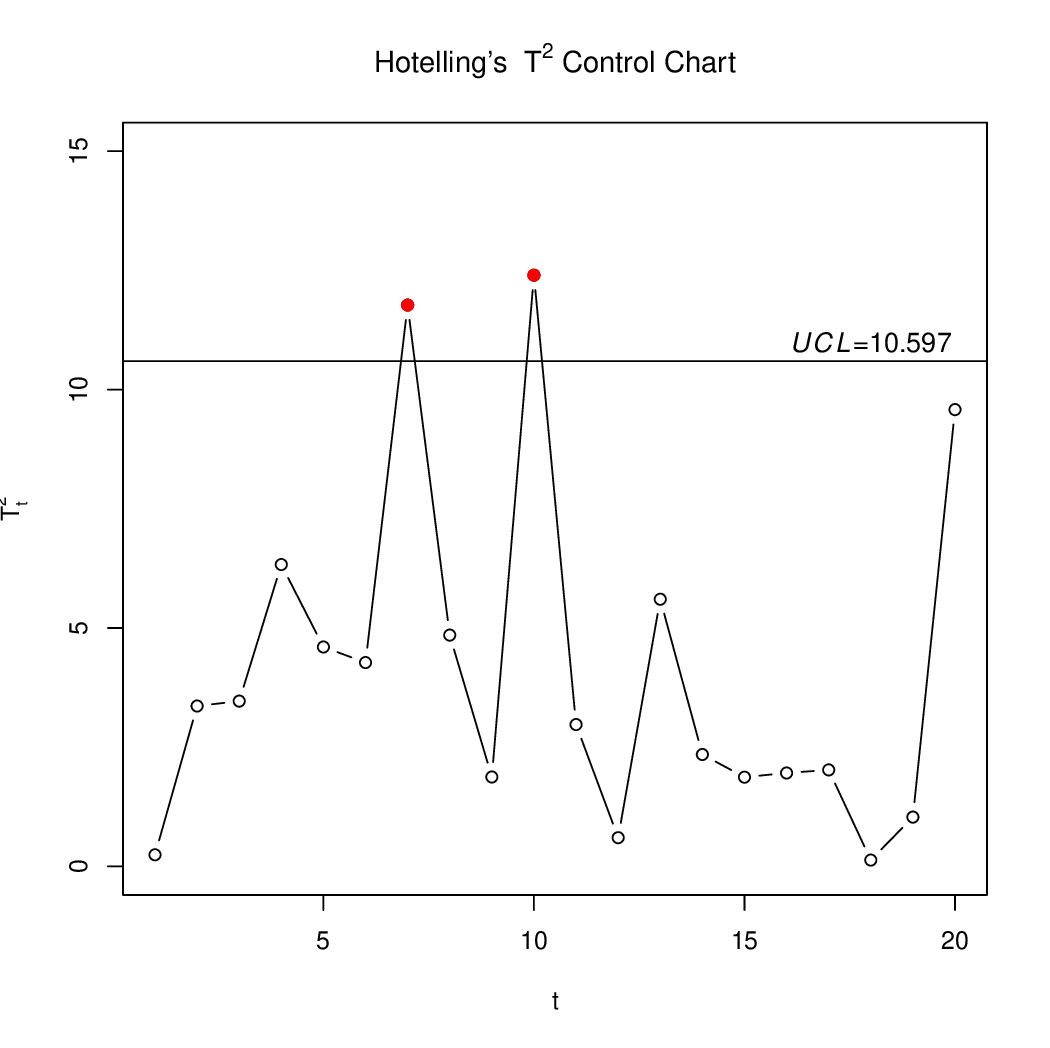}}}
 \caption{Phase I and II Hotelling's $T^2$ control chart for the chemical process example.}
\label{fig:T2chartphaseI}
\end{figure}

Figure \eqref{fig:ACFT2} (a) and (b) show the ACF and PACF plots of the monitored statistics $T_t^2$. From these figures, it can be said that 
the sample size $n=5$ is sufficient to eliminate autocorrelation between the consecutive  $T^2$ statistics. It means that even though we do not consider any time gap between consecutive samples, only an insignificant autocorrelation has been left between their corresponding monitor statistics. Since we are now in Phase I and given $ARL_0=200$, we use the F distribution quantiles to calculate the UCL of the $T^2$ chart as $UCL=\frac{v(m-1)(n-1)}{mn-m-v+1} F_{\alpha,v,mn-m-v+1}=10.910$ where $m=20$ is the total number of samples and $F_{\alpha,v,mn-m-v+1}$ is the upper $\alpha$-th percentile of the Fisher distribution with $v$ and $mn-m-v+1$ degrees of freedom. Figure \eqref{fig:T2chartphaseI} (a) shows the Hotelling's $T^2$ control chart for Phase I observations based on samples 1-20 each of size $n=5$. Since the control chart does not trigger any OC signal for Phase I observations, the estimated parameters in \eqref{equ:modelparameters} and \eqref{equ:modelparameterss}  can be used for the Phase II implementation of the control chart.

\subsubsection*{Phase II illustration}

To study the performance of the $T^2$   chart in Phase II, the process mean is artificially shifted from its IC value $\pmb{\mu}_{\textbf{X}}=(0,0)^{\intercal}$ to the OC value $\pmb{\mu}_{\textbf{X}}=(0.1,0.1)^{\intercal}$ and then 20 samples of size $n=5$ are generated from the estimated VAR(3) model. Given $ARL_0=200$, the $UCL$ of the $T^2$ control chart in Phase II is $\chi^{2}_{2}(0.005)=10.597$. Figure \eqref{fig:T2chartphaseI} (b) shows that Hotelling's $T^2$  control chart detects the shift in mean vector at inspections $\#7$ and $\#10$.

%%%%%%%%%%%%%%%%%%%%%%%%%%%%%%%%%%%%%%%%%%%%%
%%%%%%%%%%%%%%%%%%%%%%%%%%%%%%%%%
\section{Conclusions}
\label{sec:conclusions}
In this paper, we developed the methodology to implement a Hotelling $T^2$ chart for monitoring the mean of multivariate time series data following a VAR($p$) time series model. A VAR($p$) model allows us to account for both autocorrelation and cross-correlation effects among observations collected during the last $p$ sampling intervals. A comprehensive numerical study was conducted to assess the effect of various factors including the sample size, the shift size, and more importantly, the level of correlation and autocorrelation on the detection ability of the $T^2$ chart. With this study, we also discuss how the design and the statistical performance of the $T^2$ chart change with the VAR($p$) model’s parameters. The results of the study show that the association (correlation, autocorrelation, and cross-correlation) among the data can noticeably deteriorate the performance of the chart. In particular, the higher the association, the slower the chart is in detecting out-of-control conditions. We compared the statistical performance of the $T^2$ chart based on the methodology developed in this paper to the residual-based $T^2$ chart by using the average run length and the first-to-signal criteria. The results of the study show that the proposed method uniformly performs better than the existing alternative chart in terms of detecting abnormalities. The comparison study also shows that the performance of our chart is less impacted (more robust) by the association among data. While the proposed methodology is broadly applicable, some extensions to this work are of interest.

The proposed method is applicable in Phase II of process monitoring, assuming that a perfect model fit and accurate process parameter estimation have been achieved during Phase I analysis. However, to reach such an achievement, developing a monitoring technique for Phase I of a VAR($p$) process could be one possible direction for future research. On the other hand, evaluating the effect of estimation error and model misspecification in Phase I on the performance in Phase II is another avenue for further research. Because both of the approaches we compared require us to fit a model, we expect both of them could be highly affected by substantial estimation error. We designed the chart only based on statistical considerations like the in-control and out-of-control performances of the chart. However, there are many processes in which there are economical aspects like the cost of sampling and cost of inspection that should be accounted for in the design of the control chart. Thus, the economical-statistical design of the $T^2$ chart for a VAR($p$) process could be another practically relevant direction for future research. 

\section*{Disclosure of interest}
No potential conflict of interest was reported by the authors.

\section*{Data Availability Statement}

The real data set analyzed in the illustrative examples section is presented in the main text.

\section*{Funding}

No funding was received.

\end{document}